\documentclass[prd,twocolumn,showkeys]{revtex4}
\usepackage{graphicx} 
\usepackage{amssymb} 
\usepackage{amsmath}
\usepackage{epsfig}
\usepackage{slashed}   
\usepackage{psfrag}

\newcommand{\pdf}{\ensuremath{F}}

\newcommand{\jetbub}{\ensuremath{\mathcal{J}}}
\newcommand{\pdfbub}{\ensuremath{\Phi}}
\newcommand{\hardbub}{\ensuremath{\mathcal{H}}}

\newcommand{\approxa}[1]{\ensuremath{\tilde{#1}}}
\newcommand{\approxb}[1]{\ensuremath{\hat{#1}}}

\newcommand{\fieldtensor}{\ensuremath{\mathcal{G}}}

\newcommand{\strong}{\ensuremath{g_s^2 \,}}

\newcommand{\pdfamp}{\ensuremath{\mathcal{L}}}
\newcommand{\upperbub}{\ensuremath{\mathcal{U}}}
\newcommand{\trans}{\ensuremath{t}}
\newcommand{\fact}{\Xi}
\newcommand{\hadtensor}{\ensuremath{W}}
\newcommand{\picineq}[1]{ \ensuremath{\begin{array}{c} \includegraphics[scale=0.3]{#1} \end{array} } }
\newcommand\3[1]{{\bf #1}}

\begin{document}
\title{Next-to-Leading Order Hard Scattering Using Fully Unintegrated
Parton Distribution Functions} 
\author{Ted C. Rogers\footnote{Present Address: Dept. of Physics And Astronomy, Vrije Universiteit Amsterdam, 1081 HV Amsterdam, The Netherlands}}
\affiliation{Department of
Physics, Pennsylvania State University,\\ University Park, PA  16802,
USA}
\date{\today}

\begin{abstract}
We calculate the next-to-leading order fully 
unintegrated hard scattering coefficient 
for unpolarized gluon-induced deep
inelastic scattering using the 
logical framework of parton correlation functions developed in 
previous work.  In our approach, exact four-momentum conservation is 
maintained throughout the calculation.  Hence, all non-perturbative functions, like
parton distribution functions, depend on all components of parton four-momentum.
In contrast to the usual collinear factorization approach where the hard scattering coefficient 
involves generalized functions (such as Dirac $\delta$-functions), the fully unintegrated hard scattering 
coefficient is an ordinary function.
Gluon-induced deep inelastic scattering provides a simple illustration of the application of the fully unintegrated factorization formalism
with a non-trivial hard scattering coefficient, applied to a phenomenologically
interesting case.   Furthermore, the gluon-induced process 
allows for a parameterization of the fully unintegrated gluon distribution function.
\end{abstract}

\maketitle

\section{Introduction}
\label{sec:intro}
The standard factorization theorems of perturbative QCD (pQCD) are the main ingredients 
in many phenomenological calculations of high energy processes (for a review of the usual approach to collinear factorization, see~\cite{CSSreview}). 
They are especially important in Monte Carlo event generator (MCEG) calculations (see e.g.,~\cite{Skands:2005hd}) 
and, more generally, in many of the calculations relevant to up-coming experiments such as those taking place at the Large Hadron Collider. 
The standard derivations, however, 
rely on a number of approximations
that alter final state momentum such that over all four-momentum conservation 
is violated.  Hence, there is a potential for large errors when the standard
factorization theorems are extended to more general situations than what 
have been considered in the past, particularly when the details of final states 
are of interest.

To understand the origin of these issues, it is helpful to begin by briefly 
reviewing the basic structure of the more standard approaches to factorization.
To start off, we recall that 
a basic component of the standard collinear factorization theorems is the concept of a
universal parton distribution function (PDF) which describes 
the likelihood of finding a parton in the proton with a particular plus component of 
momentum. 
The partons are treated as having zero transverse and minus components of momentum in the hard scattering calculation.
In the standard collinear factorization theorems, the PDFs have precise definitions as  
expectation values of appropriate combinations of field operators~\cite{Efremov:1980kz,Collins:1981uw}.
In these definitions, there is an integration over the small transverse and minus components of parton momentum because
they are neglected in the hard scattering calculation.
Thus, the standard PDF may be referred to as the ``integrated'' PDF.
Standard collinear factorization relies directly on the approximations that allow 
the small components of four-momentum to be integrated over in the definition of the PDF (for 
a specific illustration, see the introductory sections of Ref.~\cite{CRS}).
In spite of this, the standard treatment is usually sufficient 
if one is concerned only with inclusive quantities.
It is especially important that a well-defined operator definition exists for the
PDF, since the universality of this PDF is part of what gives pQCD calculations their predictive power. 

It is now well-known that, in some situations, the transverse 
momentum of the struck parton becomes important.
Examples include the treatment of the transverse momentum distribution 
in the Drell-Yan process~\cite{Soper:1977xd}, and treatments of 
scattering at very high energies (small Bjorken-$x$) where there is 
no transverse momentum ordering of emitted gluons~\cite{BFKL}.
This has lead to the use 
of transverse momentum dependent (TMD) PDFs which have explicit dependence on transverse parton momentum. 
These are also called ``unintegrated PDFs'' (UPDFs) because the integration over transverse components of momentum
is left undone inside the definition of the PDF.
There is still an integration over the minus component of momentum, so
the factorization formula still involves approximations that violate over-all four-momentum conservation. 
However, this is not a serious problem for many applications.

An important observation is that generalizing factorization to include TMD PDFs is not a simple matter of leaving the integrals over transverse momentum 
undone in the standard definition of the PDF.
Indeed, providing consistent operator definitions for the UPDFs involves a number of complications.
One problem is that the most obvious candidate definitions suffer from divergences that arise from gluons 
with infinite rapidity in the outgoing light-like direction.
These ``light-cone'' divergences
remain even after an infrared cutoff is included (see~\cite{C03} for a review of this and related issues).  
A consistent definition, therefore, requires an explicit rapidity cutoff in the 
definition of the UPDF~\cite{Soper:1979fq,TMD}.
Furthermore, it has been shown that factorization with TMD PDFs is violated for some processes~\cite{Bomhof:2007xt,Collins:2007nk}.  
For an up-to-date review of the different types of PDFs, see Refs.~\cite{C03,Hautmann:2007gw}.  
See also Refs.~\cite{CBR,Pasquini:2008ax,Meissner:2008xs,Hautmann:2007cx,Hautmann:2007uw,Cherednikov:2007tw} for recent work
related to defining the TMD PDF.  

The standard collinear factorization formalism and formalisms that use TMD PDFs accurately describe a wide 
range of phenomena.
However, some studies~\cite{MRW1,MRW2,Hoche:2007hg,Collins:2005uv} have illustrated that, at least for calculations which require 
detailed knowledge of final states, an \emph{exact} treatment of over-all kinematics is needed.
In such an approach, the PDF should carry information about \emph{all} components of parton momentum, including both
the transverse and minus components.
The definition of the PDF in this formalism should not involve an integral over \emph{any} component of momentum.  
A PDF that depends on all components of parton four-momentum 
is therefore called ``fully unintegrated'' (or ``doubly unintegrated''). 
In this fully unintegrated treatment, the kinematics of initial and final states are
kept exact throughout the calculation.  However, as discussed above, the standard kinematic approximations are needed in the derivation of the usual collinear factorization formalism. 
Hence, as with a treatment of TMD PDFs, the fully unintegrated treatment requires a new approach to factorization.  Among other issues, a precise definition 
for the fully unintegrated PDF is needed.

A formal derivation of a fully unintegrated approach to factorization was initiated in 
Refs.~\cite{CZ,CRS}.  In~\cite{CZ}, a complete, fully unintegrated treatment of factorization was given in 
the context of an MCEG.  To avoid complications with gauge invariance, the authors restricted consideration to 
a scalar-$\phi^3$ theory in six space-time dimensions.  
There, the derivation of factorization utilized a nested subtraction scheme in which 
double counting subtractions were required to obtain the correct hard scattering coefficient at 
each order of perturbation theory.
This is a generalization of the approach to factorization discussed in~\cite{JCC} and is reminiscent of the 
Bogoliubov approach to renormalization.
Performing the subtractions consistently required the introduction in~\cite{CZ} of a well-defined 
mapping from exact to approximate parton momentum. 
The derivation in~\cite{CZ} was for arbitrarily many jets in the final state.
Though the proof was for a scalar theory only, the procedure for mapping exact to approximate parton momentum is likely 
to extend to the gauge theory case.

In~\cite{CRS}, a complete fully unintegrated derivation of factorization for a gauge theory 
was given for the case of a single out-going jet.  
The focus in Ref.~\cite{CRS} was on the leading order (LO) contribution to the hard scattering coefficient (though the 
single gluon vertex correction was also calculated).
Since the relevant Ward identities 
do not yet have explicit enough proofs for the non-Abelian case, the derivation in~\cite{CRS} only applies rigorously 
to an Abelian gauge theory. 
The resulting structure is, however, highly suggestive of a similar factorization formula for the non-Abelian case (QCD).

An important difference between the standard collinear formalism and the fully unintegrated formalism is the number of 
non-perturbative functions that appear in a given factorization formula.
In inclusive deep inelastic scattering (DIS), for example, a total cross section calculation that uses the standard collinear factorization approach only involves a single hard scattering coefficient and 
a single PDF.  Schematically, the structure of the standard collinear factorization formula is:
\begin{equation}
\sigma \sim \sum_j \mathcal{C} \otimes f_{i/p}. 
\end{equation}
Here, $\mathcal{C}$ is a hard scattering coefficient and $f_{i/p}$ is the usual integrated PDF for scattering from 
a parton of type $i$ inside the proton.  The symbol $\otimes$ denotes 
the usual convolution integral.

By contrast, the fully unintegrated approach requires not only a fully unintegrated PDF (FUPDF) in the initial state, but also fully unintegrated jet factors and soft factors to describe
final state parton four-momentum.  Thus, the fully unintegrated factorization formula found in Ref.~\cite{CRS} for scattering 
from a target quark in the proton includes not one but three non-perturbative factors: a fully unintegrated quark PDF, a fully unintegrated jet factor, and a soft factor.  
The integrals that would normally allow the final state non-perturbative factors to be simplified are left undone in the fully unintegrated treatment. 
The structure of the factorization formula in Ref.~\cite{CRS} for DIS is written schematically:
\begin{equation}
\label{factorization}
\sigma \sim \mathcal{C} \otimes F_{q/p} \otimes J \otimes S,
\end{equation} 
where $F$, $J$, and $S$ represent, respectively, the fully unintegrated quark PDF, the fully unintegrated jet factor, and the fully unintegrated soft factor.
Collectively, we refer to the fully unintegrated non-perturbative objects that appear in the fully unintegrated treatment as ``parton correlation functions'' (PCFs). 
In Ref.~\cite{CRS} the PCFs in Eq.~(\ref{factorization}) are given precise definitions as expectation values of field operators.  
Clearly, the proliferation of non-perturbative factors makes the fully unintegrated approach much more 
complicated than the usual formalism.  The advantage, however, is that it provides a much more complete description of the 
distribution of final states.

Much of the work done so far with PCFs and the fully unintegrated approach to factorization has been rather formal, with very limited phenomenological applications.
Therefore, the aim of this paper is to initiate a closer connection to phenomenology by directly applying the methods of Ref.~\cite{CZ,CRS} to a calculation of
the 
hard scattering coefficient for deep inelastic wide-angle jet pair production 
from a target gluon inside a proton.  The partonic subprocess is then,  $\gamma^{\ast} g \rightarrow q \bar{q}$ (see Fig.~\ref{fig:boxdiagrams}).   
This is the simplest possible calculation of a non-trivial next-to-leading order (NLO) hard scattering process in the fully unintegrated approach to factorization.
Since gluon-mediated processes dominate
at small values of Bjorken-$x$, the resulting hard scattering coefficient can be used 
to parameterize the PCFs. 
Conversely, with a particular model (or parameterization) of the non-perturbative factors, the fully unintegrated 
hard scattering coefficient calculated in this paper can be used to make direct predictions that are consistent with factorization.
Since the hard scattering coefficient is to be determined, as in~\cite{CZ}, from a sequence of double-counting subtractions,
the calculation relies directly on the results of~\cite{CZ,CRS}.  
The result of this paper, therefore, is a prescription that allows for a simple and direct implementation of the 
fully unintegrated approach in calculations of gluon mediated DIS.

Although the work of Refs.~\cite{CZ,CRS} is sufficient 
to provide the steps for calculating the order-$\strong$ hard scattering coefficient, 
it is worth emphasizing that a complete derivation 
of factorization for the production of two jets requires a systematic consideration of arbitrarily many collinear and soft gluons.
Additionally, the Ward identity arguments in Ref.~\cite{CRS} need to be extended to the non-Abelian case.
Since the purpose of this paper is to begin to apply the results of 
Refs.~\cite{CZ,CRS} to phenomenology, we leave these important but more formal issues to future work.
Additionally, we set aside for now the issue of the evolution of the PCFs, initial/final state showering, or the relationship 
to other integrated or unintegrated gluon distribution functions.  

The paper is structured as follows: In Sect.~\ref{sec:setup}, we establish conventions and 
set up notation.  In Sect.~\ref{sec:strategy}, we outline the basic steps in the calculation of 
a fully unintegrated hard scattering coefficient.  In Sect.~\ref{sec:lowangle}, we go through the steps for low-angle scattering
and write down the LO factorization formula. 
in Sect.~\ref{sec:hardregion} we discuss scattering with large transverse momentum. 
In Sect.~\ref{sec:double} we go through the steps for obtaining the subtraction terms that 
allow for a correct evaluation of the wide-angle (high transverse momentum) hard scattering coefficient.
In Sect.~\ref{sec:NLO}, we write down the NLO factorization formula. 
In Sect.~\ref{eq:complete} we summarize the results and make concluding remarks.
 
\section{Set-up and Notation}
\label{sec:setup}

\begin{figure*}
\centering
  \begin{tabular}{c@{\hspace*{5mm}}c}
    \includegraphics[scale=0.4]{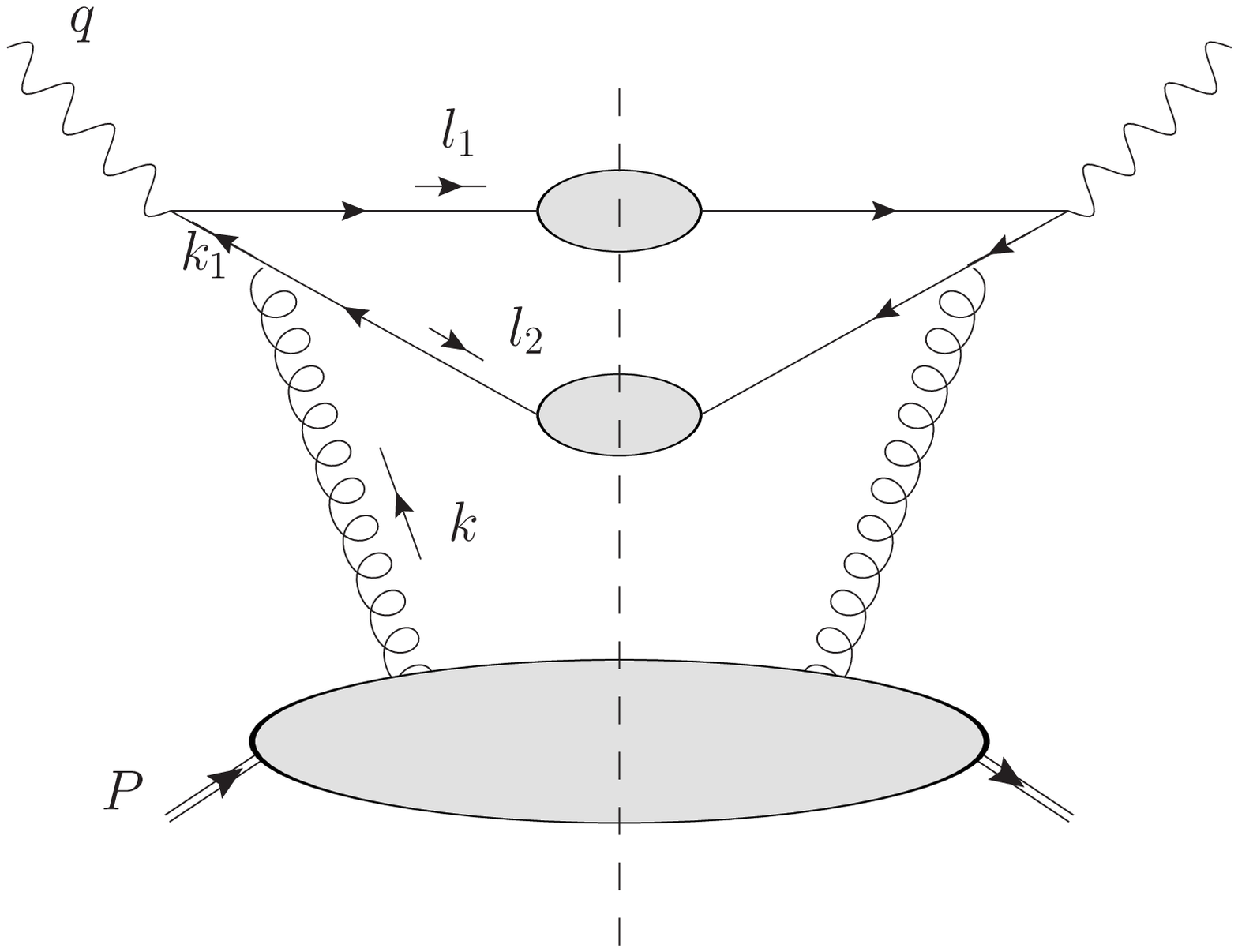}
    &
    \includegraphics[scale=0.4]{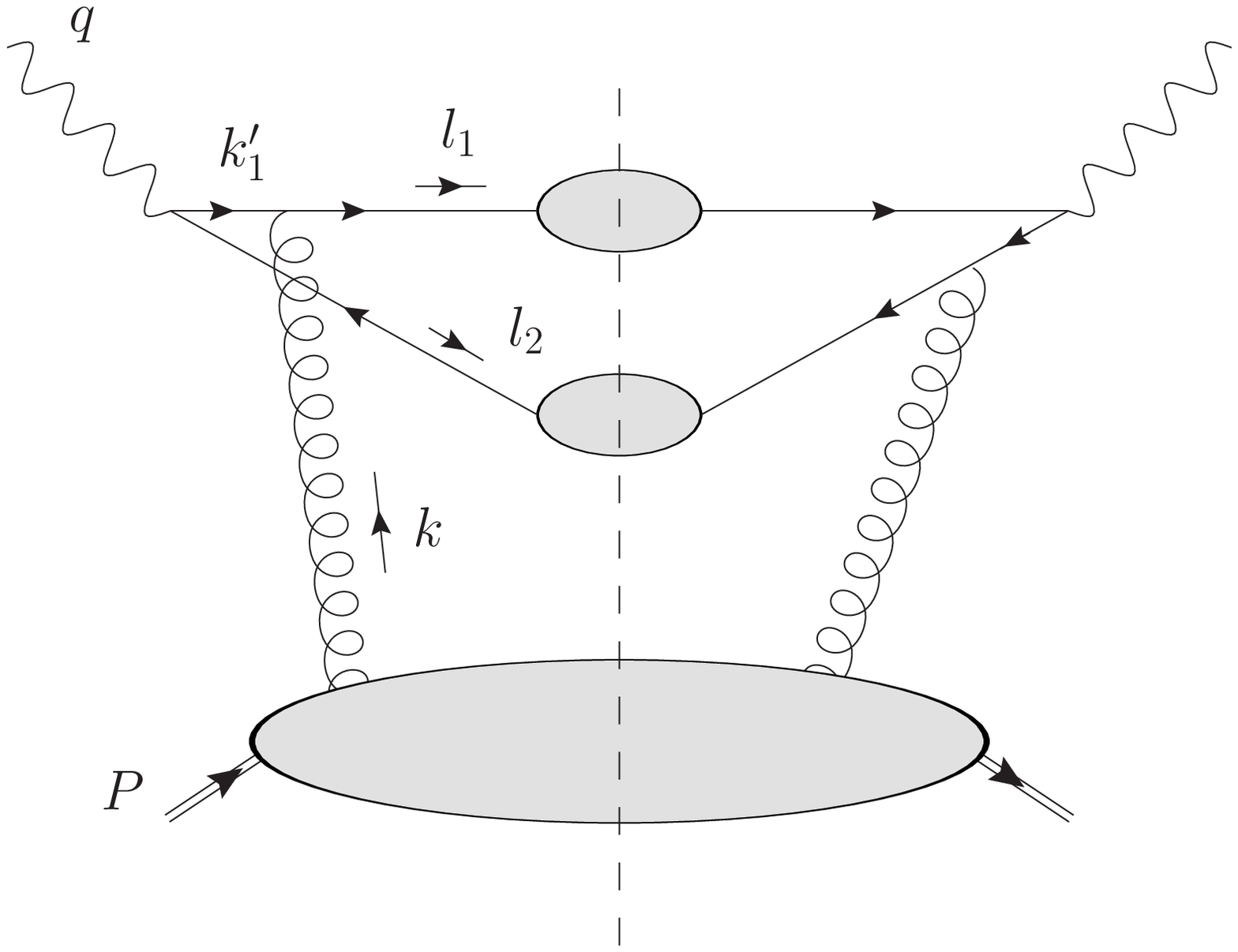}
  \\
    $(a)$ & $(b)$ \\
  \includegraphics[scale=0.4]{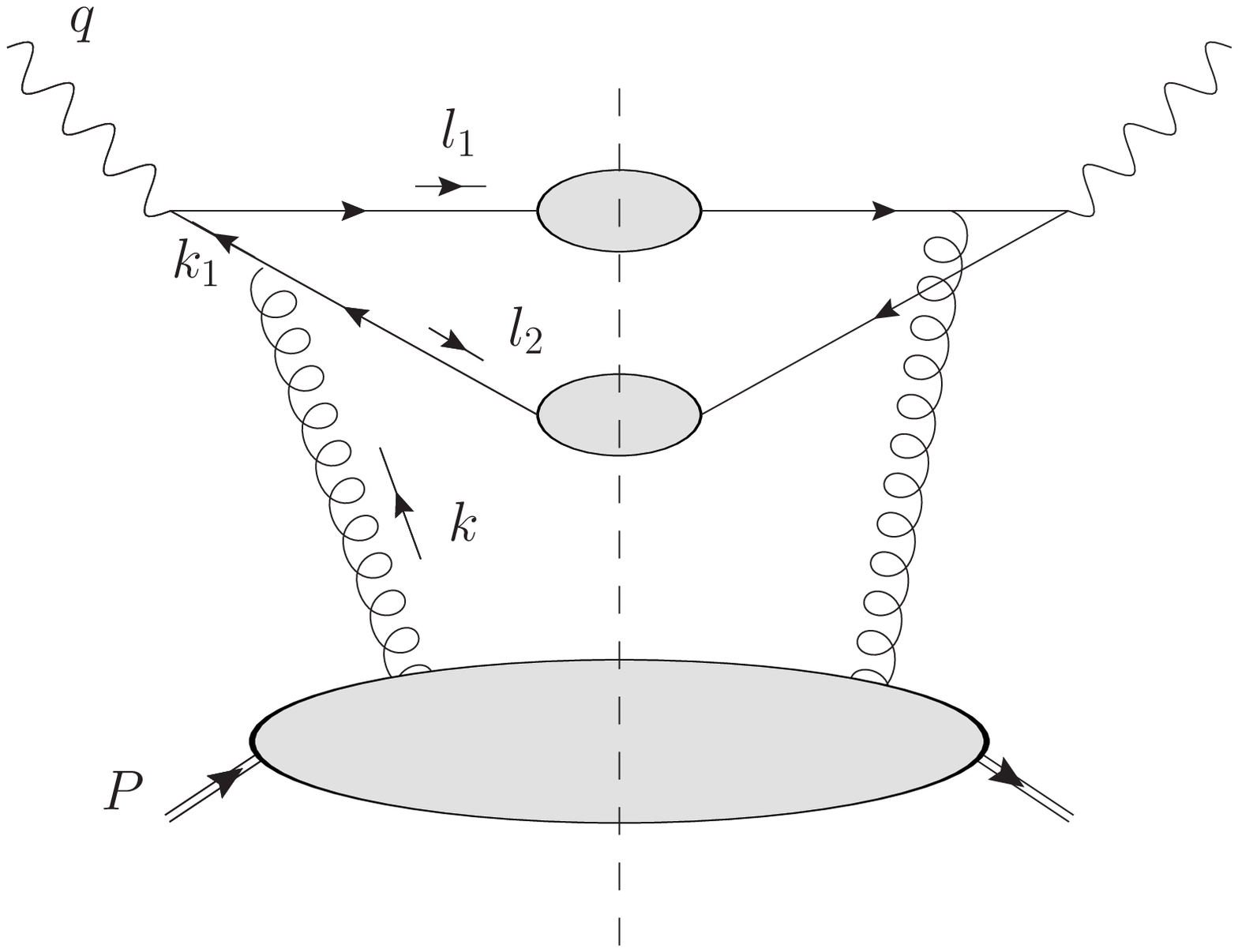}
    &
    \includegraphics[scale=0.4]{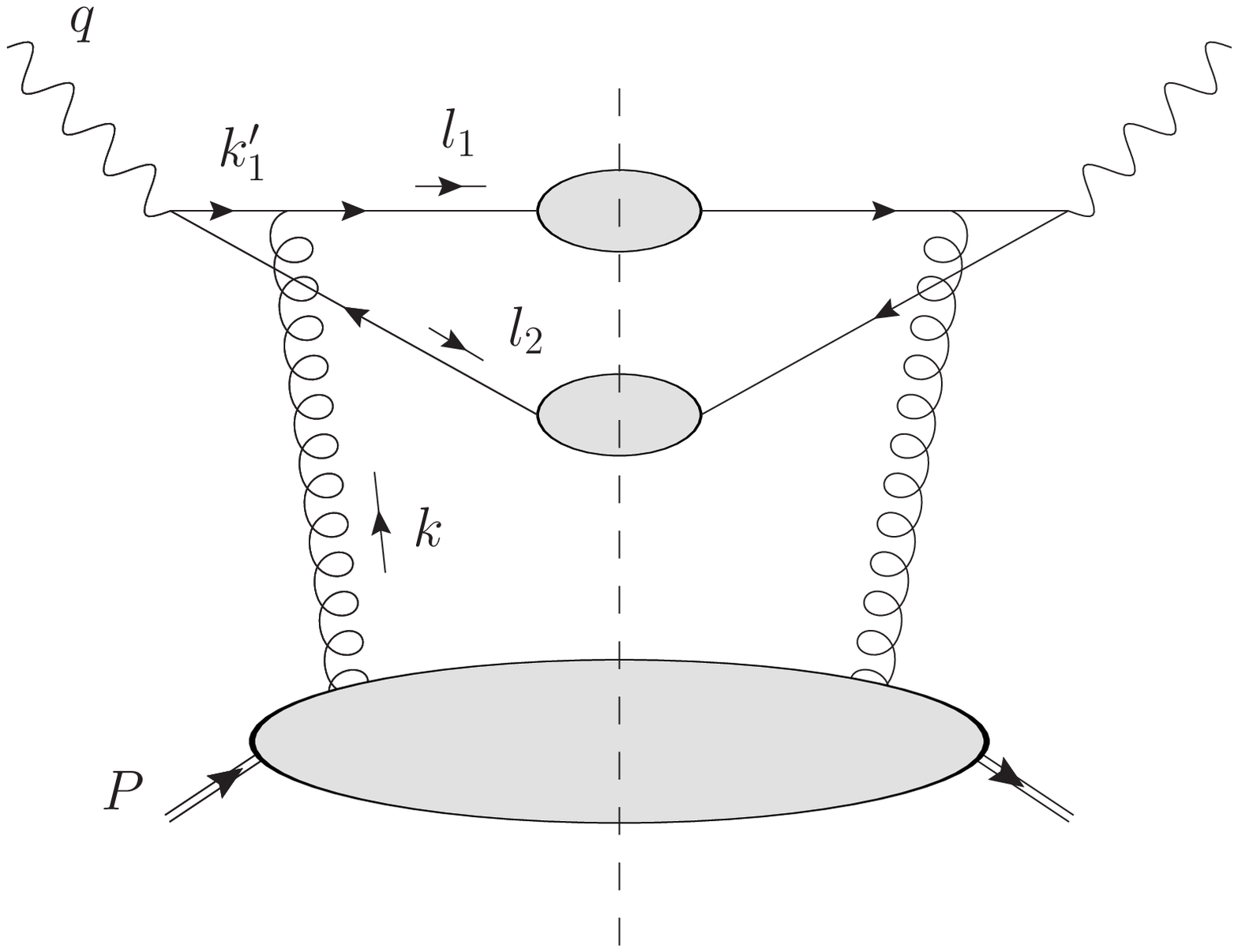}
  \\
    $(c)$ & $(d)$ 
  \end{tabular}
\caption{Graphs contributing to gluon induced production of two jets at order-$\strong$.}
\label{fig:boxdiagrams}
\end{figure*}

The process under consideration is described by 
the Feynman diagrams shown in Fig.~\ref{fig:boxdiagrams}.
The bubbles represent target and final state subgraphs.
All calculations are done in Feynman gauge.
In a more precise treatment, we would also need to
include a final state soft factor, but this 
will not be critical at the order of perturbation theory that we consider here.  (If we continue 
to higher orders, a soft factor needs to be included, as in Ref.~\cite{CRS}.)
We will usually express four-vectors in terms of light-cone variables.
If $V=(V^+,V^-,{\bf V}_t)$ is a four-vector, then we use the convention that
\begin{eqnarray}
V^+ = \frac{V^0 + V^3}{\sqrt{2}},\\
V^- = \frac{V^0 - V^3}{\sqrt{2}},\\
{\bf V}_t = (V^1,V^2).
\end{eqnarray}
Then
\begin{equation}
V^2 = 2 V^+ V^- - V_t^2.
\end{equation}

Following standard conventions, we work in a frame where the total 
transverse momentum of the photon-proton collision is zero, and express the incoming proton and virtual photon as
\begin{equation}
P = \left( P^+, \frac{M_p^2}{2 P^+}, {\bf 0}_t  \right), \qquad q = 
\left( -x P^+, \frac{Q^2}{2 x P^+}, {\bf 0}_t \right).
\end{equation}
Since we insist on using exact kinematics throughout, $x$ is not the usual Bjorken-x variable but 
rather
\begin{equation}
x = \frac{2 x_{Bj}}{1 + \sqrt{1 + 4 \frac{M_p^2}{Q^2} x_{Bj}^2 } },
\end{equation}
with $x_{Bj}$ being the usual Bjorken-x variable:
\begin{equation}
x_{Bj} = \frac{Q^2}{2 P \cdot q}.
\end{equation}
Additionally, it will be convenient to work in the Breit frame where $x P^+ = Q/\sqrt{2}$.
The momenta of the final state jets are $l_1$  and $l_2$, and their masses are $M_1^2$ and $M_2^2$ respectively.
The target gluon is labeled by $k$.  Momentum labels are shown in Fig.~\ref{fig:boxdiagrams}.

The contribution to the hadronic structure tensor from the graphs in Fig.~\ref{fig:boxdiagrams} is given by the expression
\begin{multline}
\label{eq:gammasummod}
\hadtensor_g^{\mu \nu}(P,q) = \\ \frac{e_j^2}{4 \pi} \int \frac{d^4 l_{2}}{(2 \pi)^4 } 
\int \frac{d^4 l_{1}}{(2 \pi)^4 } \int \frac{d^4 k}{(2 \pi)^4 } 
\left( \left| M(l_1,l_2,k) \right|^2 \right)^{\mu \nu}  \times \\ \times (2 \pi)^4 \delta^{(4)}(k + q - l_1 - l_2).
\end{multline}
We use the letter $M$ to denote amplitudes for the full process: $\gamma^{\ast} p \rightarrow 2 \,\, {\rm Jets} + X$, in 
contrast to $A$ which will be reserved to denote amplitudes for partonic subprocesses.
The subscript $g$ on the hadronic tensor symbolizes that, in the present treatment, we only consider graphs with a gluon in the target.
In our analysis, it will be convenient to treat the final state jet momenta as independent.
Therefore, we leave the integral over the momentum-conserving $\delta$-function in Eq.~(\ref{eq:gammasummod}) undone 
until the very end. 
Throughout all calculations in this paper we use the convention that a factor of $e_j^2/4 \pi$ is pulled out front as in Eq.~(\ref{eq:gammasummod}).  
The subscript $j$ labels quark flavors, and any sum over flavors is assumed to be implicit.
A structure function is obtained from Eq.~(\ref{eq:gammasummod}) by making appropriate projections onto the electromagnetic vertices.
For example, for the $F_1$ structure function we have
\begin{equation}
\label{eq:F1}
F_1(P,q) = P_{\mu \nu} \hadtensor^{\mu \nu}(P,q),
\end{equation} 
where
\begin{equation}
\label{eq:F1proj}
P_{\mu\nu} 
= \frac12 
\left[ -g_{\mu\nu} + \frac{ Q^2 P_\mu P_\nu }{ (P\cdot q)^2 +M_{p}^2 Q^2 } \right].
\end{equation}

\section{Strategy}
\label{sec:strategy}

To calculate the hard scattering coefficients in Fig.~\ref{fig:boxdiagrams} in a way 
that is consistent with the methods of~\cite{CZ,CRS} we take the following steps:
\begin{enumerate}
\item {\bf For each graph, categorize the leading regions of momentum space.}\\
\linebreak
The leading regions in momentum space for a given Feynman graph can be categorized according to 
the pinch-singularity surfaces (PSSs) of the Libby and Sterman analysis~\cite{Libby:1978bx}.
The two leading regions contributing in Fig.~\ref{fig:boxdiagrams} are: 
the target collinear region, $R_1$, where the struck quark ($k_1 = k - l_2$ in graph (a))  
is collinear to the target proton; and the hard (wide-angle) region, $R_2$,
where the outgoing final state jets are at wide angles relative to 
each another with large transverse momenta.  
There is also a leading region, $\bar{R}_1$, 
where the target-collinear parton is a struck anti-quark with momentum $k_1^\prime = l_1 - k$ (see graph (d)).
We will always assume that the gluon momentum, $k$, is 
collinear to the target proton: 
\begin{equation}
\label{eq:ksize}
k \sim \left(Q,\frac{\Lambda^2}{Q}, \Lambda \right).
\end{equation}
Corrections for gluon momenta away from Eq.~(\ref{eq:ksize}) are to be dealt with in higher order calculations.
(For an elementary discussion of categorizing leading regions in terms of PSSs see, e.g., Ref.~\cite{Sterman:1994ce}.)
\item 
{\bf For each leading region, apply approximations on 
lines entering the hard scattering subgraph, appropriate to that particular leading region.}\\
\linebreak
In the scalar theory this only involves replacing exact kinematic
variables by approximate ones \emph{inside the hard scattering subgraph}.
Here it is important to note that initial and final state momenta are not altered at all in the approximation.

A well-defined mapping from exact to approximate variables is required 
to ensure consistency with factorization.
Such a mapping for arbitrarily many outgoing jets was given in Ref.~\cite{CZ} for a scalar theory.
In a gauge theory, one also needs to project appropriate external parton polarizations, as 
is done for the external quarks in Ref.~\cite{CRS}.  
In this paper, we will need analogous  
projections to extract appropriate gluon polarizations.
Using the notation of~\cite{CRS}, the complete set of replacements in the 
hard scattering subgraph, including polarization projections, will be symbolized by the action 
of an ``approximator'' on the unapproximated expression, Eq.~(\ref{eq:gammasummod}).
Since a different set of approximations is appropriate for each region, then each region is associated with its own approximator.
For example, $T_1$ symbolizes the approximator appropriate to region $R_1$, and 
$T_1 \hadtensor^{\mu \nu}(P,q)$ is a good approximation to Eq.~(\ref{eq:gammasummod}) in region $R_1$.
So, $T_1 \hadtensor^{\mu \nu}(P,q)$ is to be read as an instruction to replace the exact momenta 
in the integrand of Eq.~(\ref{eq:gammasummod}) with approximate momenta appropriate for region $R_1$, and to 
make appropriate projections on parton polarizations.
Analogous approximators, $\bar{T}_1$ and $T_2$, are associated with regions $\bar{R}_1$ and $R_2$.
(Each of these approximations/approximators will be defined explicitly in later sections.)
In each case, the approximator makes substitutions that are good approximations in the corresponding leading region.
This notation will make the logic of the subtractions straightforward.  It is important for this logic that, although each approximation 
is only good in a certain region of momentum space, each is well-defined for all momenta.  
\item  
{\bf In each leading region, use Ward identities to disentangle soft and collinear 
gluons into factors that are identifiable as contributions to PCFs.} \\
\linebreak
In the calculation of this paper, a Ward identity argument will allow the gluon to be disentangled from the out-going quark
when the struck quark or anti-quark is collinear to the target proton (regions $R_1$ and $\bar{R}_1$).
In the resulting factorized structure, one of the factors will be identified as 
the zeroth order contribution to the fully unintegrated hard scattering coefficient.
The other factor will be identified as an order-$\strong$ contribution
to a perturbative expansion of the fully unintegrated quark or anti-quark PDF.  
The factor corresponding to a contribution to the fully unintegrated PDF can then be seen to be 
consistent with 
the operator definition for the fully unintegrated quark PDF given in Ref.~\cite{CRS}.

For a detailed discussion of the PCFs and their definitions, see 
section V of Ref.~\cite{CRS}.  Here we very briefly review the definition given there 
for the fully unintegrated quark PDF.

In Ref.~\cite{CRS}, the fully unintegrated quark PDF was defined in coordinate space to be
\begin{multline}
\; \; \; \; \; \; \tilde{F}(w,y_p,y_s,\mu)
\\
=
\langle P | \bar{\psi}(w) V^{\dagger}_{w}(n_s)  I_{n_s;w,0}  
           \dfrac{\gamma^{+}}{2}
V_{0}(n_s) \psi(0) | P \rangle_{R}.
\label{quarkpcf}
\end{multline}
The momentum space distribution is then found by Fourier transforming.
Let us briefly describe each of the elements of Eq.~(\ref{quarkpcf}):
$\psi$ is the quark field operator, and  $| P \rangle$ is the target proton state.
Wilson line operators, $V$,$V^{\dagger}$,$I$, are needed in Eq.~(\ref{quarkpcf}) to ensure exact gauge invariance. 
Hence, we define
$V_{w}(n_s)$ as the path-ordered exponential of gauge fields from 
coordinate $w$ to $\infty$:
\begin{multline}
\label{eq:WL}
\; \; \; \; \; \; V_w(n) = \\ {\rm P} \exp \!\left( -ig_s \int_0^{\infty} d\lambda \, \,
                    n \cdot A(w+\lambda n) \right) \; ,
\end{multline}
where $A$ is the gauge field and ${\rm P}$ is a path-ordering operator.
The Wilson line direction is nearly at rest in the center of mass system
\begin{equation}
\label{eq:ns}
n_s = (-e^{y_s},e^{-y_s},\3{0}_t) \qquad y_s \sim 0.
\end{equation}
This direction is chosen to 
give the most universal derivation of factorization~\cite{CM}.
Furthermore,
the fact that the Wilson line direction is nonlightlike ensures that light-cone divergences associated with 
gluons moving with infinite negative rapidity are regulated~\cite{Soper:1979fq,TMD}.
In an extension to the non-Abelian gauge theory, the exponent in Eq.~(\ref{eq:WL}) will also need
to include the generator for the gauge group in the fundamental representation. 

For exact gauge invariance, Eq.~(\ref{quarkpcf}) also needs a Wilson line segment, $I_{n_s;w,0}$, at light-cone 
infinity to connect the other two lines~\cite{Belitsky:2002sm}.  The definition of this gauge link at infinity is,
\begin{equation}
\label{pdfinf}
I_{n,w,0} = {\rm P} \exp \!\left( -ig_s \int_C dz^{\mu} A_{\mu}(z) \right) \; ,
\end{equation}
where $C$ is a contour that connects the points at infinity.  This operator yields a factor
of unity for our calculations since we work in Feynman gauge.  The combination of Wilson line 
operators in Eq.~(\ref{quarkpcf}) forms a continuous Wilson line that connects point $0$ to point $w$.

In Eq.~(\ref{quarkpcf}), $\mu$ is the renormalization scale associated with the usual ultra-violet divergences.
The subscript $R$ on the right side of Eq.~(\ref{quarkpcf}) signifies that renormalized operators are used.  
Thus, the fully unintegrated quark PDF depends on the quark momentum, $k$, the proton rapidity $y_p$, 
the rapidity parameter $y_s$ used to parameterize the Wilson line direction, and the renormalization 
scale $\mu$.  For a detailed discussion of these issues and the motivation for Eq.~(\ref{quarkpcf}) as the 
definition of the fully unintegrated quark PDF, see section VC of Ref.~\cite{CRS}.
\item {\bf Perform all necessary double counting subtractions.} \\
\linebreak
The complete factorized expression for Eq.~(\ref{eq:gammasummod}) is obtained via 
subtractions as in~\cite{JCC,CZ}.  We briefly summarize the steps here.

In a first approximation, contributions from high transverse momentum jets 
can be neglected because they are suppressed by powers of $\strong$ relative 
to the LO contribution while for low transverse momentum we cannot rely on asymptotic freedom.
If this level of accuracy were sufficient,
then we would only need to analyze regions $R_1$ and $\bar{R}_1$, which were already considered in detail 
in~\cite{CRS}.  That is, we could use the approximator notation from step 2 to write
\begin{multline}
\label{eq:lowangle}
\; \; \; \; \; \; \hadtensor_g^{\mu \nu}(P,q) = T_1 \hadtensor_g^{\mu \nu}(P,q) + \bar{T}_1 \hadtensor_g^{\mu \nu}(P,q) + \\ +
\mathcal{O}\left( \left( \frac{\Lambda}{Q}  \right)^a | \hadtensor_g^{\mu \nu}(P,q) | \right) + \mathcal{O}\left( \strong \right),
\end{multline}
where $a > 0$ and $\Lambda$ is a typical hadronic mass scale.    
We symbolize the contributions from regions $R_1$ and $\bar{R}_1$ by
\begin{eqnarray}
\hadtensor^{\mu \nu}_{\gamma^{\ast} q \to q}(P,q) = T_1 \hadtensor_g^{\mu \nu}(P,q), \\ \bar{\hadtensor}^{\mu \nu}_{\gamma^{\ast} \bar{q} \to \bar{q}}(P,q) = \bar{T}_1 \hadtensor_g^{\mu \nu}(P,q).
\end{eqnarray}
We know from step 3 above that these contributions to the hadronic tensor 
will ultimately be expressed in terms of order-$g_s^2$ contributions to the fully unintegrated quark (anti-quark) PDFs.
  
To include the contribution from large transverse momentum scattering, we first recognize that the remainder terms in Eq.~(\ref{eq:lowangle}) are
given exactly by the subtracted expression,
\begin{equation}
\label{eq:subtracted}
{\rm remainder} = (1 - T_1 - \bar{T}_1) \hadtensor_g^{\mu \nu}(P,q). 
\end{equation}
Up to order-$g_s^4$ corrections, this remainder term is dominated by wide-angle jets originating from 
the produced quark and antiquark, i.e., from the large transverse momentum region, $R_2$, in the graphs of Fig.~\ref{fig:boxdiagrams}.
In approximator notation, $T_2$ implements approximations that are good in this region.
Thus, we apply $T_2$ to Eq.~(\ref{eq:subtracted}) to get a good approximation to the contribution from region $R_2$:
\begin{equation}
\label{eq:subs}
\; \; \; \; \; \; \; \; \; \; \; \hadtensor^{\mu \nu}_{\gamma^{\ast} g \to q \bar{q}}(P,q) = T_2 (1 - T_1 - \bar{T}_1) \hadtensor_g^{\mu \nu}(P,q).
\end{equation}
The subscript, $\gamma^{\ast} g \to q \bar{q}$, indicates the relevant hard partonic subprocess for region $R_2$.
Instead of Eq.~(\ref{eq:lowangle}) we then have,
\begin{widetext}
\begin{multline}
\label{eq:wideangle}
\hadtensor_g^{\mu \nu}(P,q) = (T_1 + \bar{T}_1) \hadtensor_g^{\mu \nu}(P,q) + (1 - T_1 - \bar{T}_1) \hadtensor_g^{\mu \nu}(P,q) \\ = (T_1 + \bar{T}_1) \hadtensor_g^{\mu \nu}(P,q) + T_2 (1 - T_1 - \bar{T}_1) \hadtensor_g^{\mu \nu}(P,q) +
\mathcal{O}\left( \left( \frac{\Lambda}{Q}  \right)^a | \hadtensor_g^{\mu \nu}(P,q) | \right) + \mathcal{O}\left( g_s^4 \right) \\
 \hadtensor^{\mu \nu}_{\gamma^{\ast} q \to q}(P,q) + \bar{\hadtensor}^{\mu \nu}_{\gamma^{\ast} \bar{q} \to \bar{q}}(P,q) + \hadtensor^{\mu \nu}_{\gamma^{\ast} g \to q \bar{q}}(P,q) +
\mathcal{O}\left( \left( \frac{\Lambda}{Q}  \right)^a | \hadtensor_g^{\mu \nu}(P,q) | \right) + \mathcal{O}\left( g_s^4 \right).
\end{multline}
\end{widetext}
A demonstration that the errors in Eq.~(\ref{eq:wideangle}) are actually  
$\Lambda/Q$-suppressed appears in Ref.~\cite{CZ} (for scalar field theory).

The first term in the last line of Eq.~(\ref{eq:wideangle}) is the contribution to the structure tensor
from the reaction, $\gamma^\ast p \to J_q + X$, i.e., the production of a single jet 
due to the hadronization of a single knocked-out quark.
The second term is the analogous contribution for the case where the jet originates 
from a knocked-out anti-quark.
Both of these terms are examples of the zeroth order contributions
to DIS that were already treated in Ref.~\cite{CRS}.

The third term in Eq.~(\ref{eq:wideangle}) is the contribution to the 
hadronic structure tensor from the reaction, $\gamma^\ast p \to J_q + J_{\bar{q}} + X$, i.e., 
from the production of two wide-angle jets that result from the hadronization of a knocked-out
quark and anti-quark.  The corresponding hard scattering coefficient will be of order $\strong$.
The last two terms represent the suppressed remainder terms.  

The first three terms in the last line of Eq.~(\ref{eq:wideangle}) give the contribution from the graphs
in Fig.~\ref{fig:boxdiagrams} to gluon-induced DIS, up to order-$g_s^4$ and power suppressed corrections.
The first two terms will factorize into a zeroth order hard scattering coefficient,
a fully unintegrated quark PDF, and a fully unintegrated jet factor, just as in Ref.~\cite{CRS}.
The subtractions in Eq.~(\ref{eq:subs}) will then allow the third term to be written in an analogous factorized form.
The aim of this paper is to find explicit expressions 
for each of the first three terms in Eq.~(\ref{eq:wideangle}) in terms of fully unintegrated PDFs and 
fully unintegrated jet factors.

\end{enumerate}

In the rest of this paper, we will go through the details of the steps enumerated above, ultimately 
arriving at an explicit expression for the factorization formula in Eq.~(\ref{eq:wideangle}).

\section{Region $R_1$: Target Collinear Region}
\label{sec:lowangle}

\subsection{Figure~\ref{fig:boxdiagrams}(a)}
We begin by considering the target collinear region $R_1$, starting with Fig.~\ref{fig:boxdiagrams}(a).
The steps followed in this section are essentially the same as those
described in Ref.~\cite{CRS}, but illustrated in detail for the specific 
case of the graphs shown in Fig.~\ref{fig:boxdiagrams}.
Since the quark is target collinear, we will apply approximations 
that allow the unapproximated graphs to be factored into 
a zeroth order hard scattering coefficient, and a factor identifiable 
as an order-$\strong$ contribution to an expansion of the fully unintegrated quark PDF.
We restrict the detailed discussion here to the case of a struck quark because
the treatment of the anti-quark case follows analogous steps.

\begin{figure*}
\centering
    \includegraphics[scale=0.6]{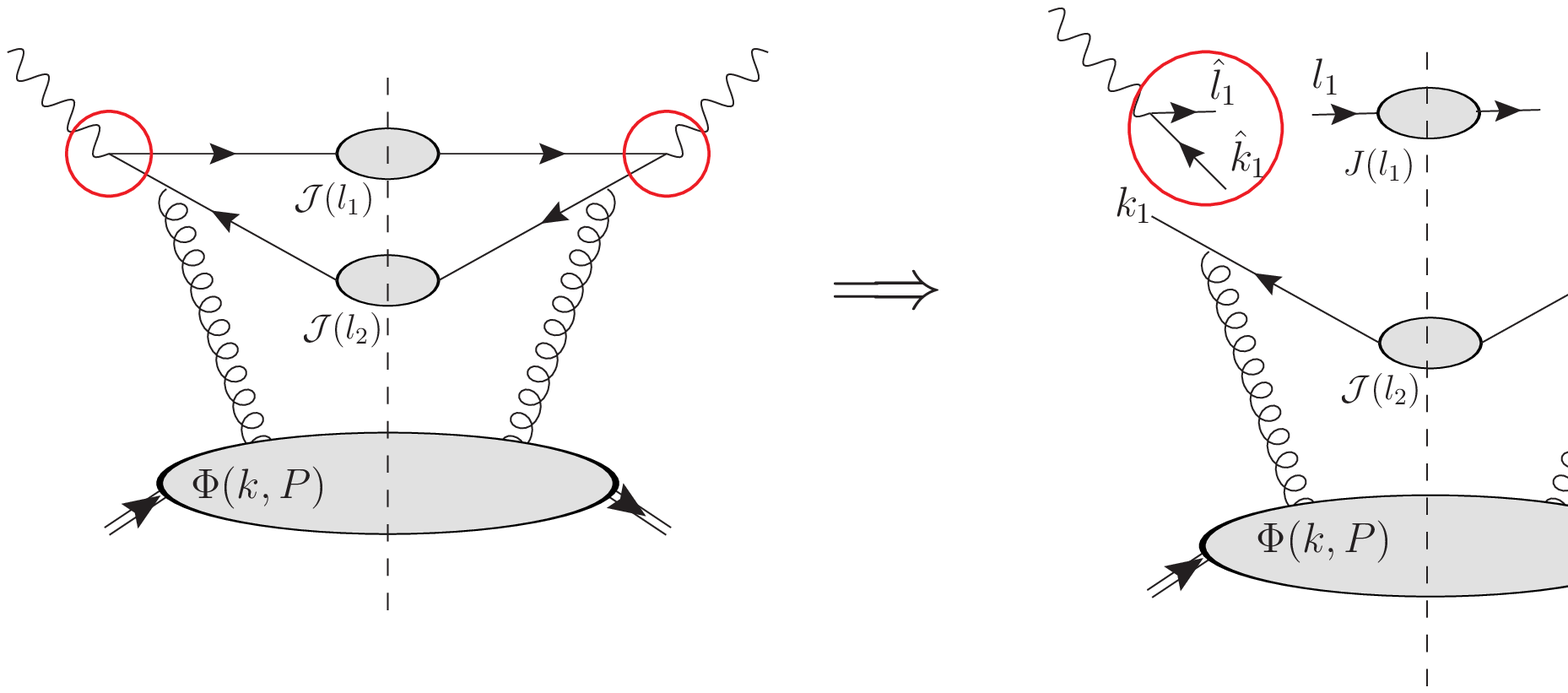}
\caption{Figure~\ref{fig:boxdiagrams}(a) with the application of the target collinear approximation for region $R_1$.  
The graph on the right-hand side of the arrow shows the separation into the factors 
of Eq.~(\ref{eq:grapha_app}).}
\label{fig:grapha_app}
\end{figure*}

The components of the struck quark momentum in region $R_1$ are of order
\begin{equation}
\label{eq:regionR1}
k_1 \sim \left( Q, \frac{\Lambda^2}{Q}, \Lambda  \right).
\end{equation}
Recall that
\begin{equation}
\label{eq:k1def}
k_1 = k - l_2.
\end{equation}
We define approximate momentum variables appropriate 
for $R_1$ with a ``hat,''
\begin{equation}
\label{eq:core}
\approxb{k}_1 = (-q^+,0,\3{0}_t), \qquad \approxb{l}_1 = (0,q^-,\3{0}_t ).
\end{equation}
These are exactly the values of the parton momentum obtained in the parton model.
Now we set up a systematic symbolic/diagrammatic notation to describe 
the approximator, $T_1$.
For region $R_1$ we should 
treat the electromagnetic vertex as the basic partonic interaction ---
it should be regarded as a separate unit which (after approximation)
can be analyzed independently of the rest of the graph.
Therefore, we graphically
denote the target-collinear approximation by inserting a solid red circle around 
the electromagnetic vertex as shown in Fig.~\ref{fig:grapha_app}.  The circle should be read as
an instruction to replace momenta passing inside it by approximate variables
appropriate for $R_1$.  Additionally, at each intersection of a propagator
line with the circle, there is a projection matrix needed to project out the 
relevant Dirac components for unpolarized scattering.   
We call the lowest order hard scattering subgraphs on the left and right sides of graph (a) $\hardbub^{L,R}_{LO,a}(l_1,k_1)$. 
(They are rather trivial at this order since they only involve the electromagnetic vertex factor $\gamma^{\mu}$.)
The appropriate target-collinear approximation (as determined in~\cite{CRS}) is to make the replacement
\begin{eqnarray}
\mathcal{H}^L_{LO,a}(l_1,k_1) \to \mathcal{P}_{\rm T} 
\mathcal{H}^L_{LO,a}(\approxb{l}_1,\approxb{k}_1) \mathcal{P}_{\rm T}, \\ \mathcal{H}^R_{LO,a}(l_1,k_1) \to \mathcal{P}_{\rm J}
\mathcal{H}^R_{LO,a}(\approxb{l}_1,\approxb{k}_1) \mathcal{P}_{\rm J}.
\end{eqnarray}
The placement of the projection operators, $\mathcal{P}_{\rm T}$ and $\mathcal{P}_{\rm J}$, is 
represented in Fig.~\ref{fig:grapha_app}a by the points where the parton lines intersect the red circles.
The projection operators are the same as in~\cite{CRS} and can be conveniently expressed in 
terms of the approximate parton momenta:
\begin{equation}
\label{eq:pmproj}
\mathcal{P}_{\rm T} = \frac{1}{Q^2} \slashed{\approxb{k}}_1 \slashed{\approxb{l}}_1, \qquad
\mathcal{P}_{\rm J} = \frac{1}{Q^2} \slashed{\approxb{l}}_1 \slashed{\approxb{k}}_1.   
\end{equation}
Explicitly, the 
unapproximated squared amplitude averaged over color for graph (a) is
\begin{widetext}
\begin{equation}
\label{eq:boxdigrama}
\left( \left| M_{a}(l_1,l_2,k) \right|^2 \right)^{\mu \nu} = \strong T_R {\rm Tr} \left[ \gamma^\nu 
\jetbub(l_1) \gamma^\mu \left( \frac{1}{\slashed{k}_1 - m} \right) \gamma^{\kappa} \jetbub(l_2) 
\gamma^\rho \left( \frac{1}{\slashed{k}_1 - m} \right) \right] \Phi_{\kappa \rho}(k,P),
\end{equation}
where the color factor is $T_R = 1/2$ for an $SU(3)$ gauge theory and $m$ is the quark mass.  
The factors $\Phi_{\kappa \rho}(k,P)$ and $\jetbub(l_1)$ and $\jetbub(l_2)$ represent, respectively,  
the incoming target subgraph and the final state jet subgraphs.
These are represented by the bubbles in Fig.~\ref{fig:grapha_app}.
The red circles in Fig.~\ref{fig:grapha_app} are to be read as an instruction to make the replacements
\begin{equation}
\label{eq:proj1}
\gamma^{\mu} \to \mathcal{P}_J \gamma^\mu \mathcal{P}_J, \qquad
\gamma^{\nu} \to \mathcal{P}_T \gamma^\nu \mathcal{P}_T,
\end{equation}
after which the amplitude becomes,
\begin{align}
T_1 \left( \left| M_{a}(l_1,l_2,k) \right|^2 \right)^{\mu \nu} =  \frac{\strong T_R}{(2 Q^2 )^2 } {\rm Tr} \left[ \gamma^\nu 
\slashed{\approxb{l}}_1 \gamma^\mu \slashed{\approxb{k}}_1 \right] {\rm Tr} \left[ \slashed{\approxb{k}}_1 \jetbub(l_1) \right] 
{\rm Tr} \left[ \slashed{\approxb{l}}_1  \left( \frac{1}{\slashed{k}_1 - m} \right)  \gamma^{\kappa} \jetbub(l_2) 
\gamma^\rho \left( \frac{1}{\slashed{k}_1 - m} \right) \right] \Phi_{\kappa \rho}(k,P) \nonumber \\
= \frac{1}{Q^2 } \left( \overline{\left| A^{\gamma^{\ast} q \rightarrow q}
(\approxb{l}_1,\approxb{k}_1) \right|}^2 \right)^{\mu \nu} \left\{ {\rm Tr}  \left[ \gamma^- \jetbub(l_1) \right] \right\}
\left\{ \strong T_R  {\rm Tr} \left[\frac{\gamma^+}{4} \left( \frac{1}{\slashed{k}_1 - m} \right) \gamma^{\kappa} \jetbub(l_2) 
\gamma^\rho \left( \frac{1}{\slashed{k}_1 - m} \right) \right] \Phi_{\kappa \rho}(k,P) \right\},
\label{eq:grapha_app}
\end{align}
where
\begin{equation}
\label{eq:vertex}
\left( \overline{\left| A^{\gamma^{\ast} q \rightarrow q}
(\approxb{l}_1,\approxb{k}_1) \right|}^2 \right)^{\mu \nu} = 
\frac{1}{2} {\rm Tr} \left[ \gamma^\nu 
\slashed{\approxb{l}}_1 \gamma^\mu \slashed{\approxb{k}}_1 \right] = \left| \picineq{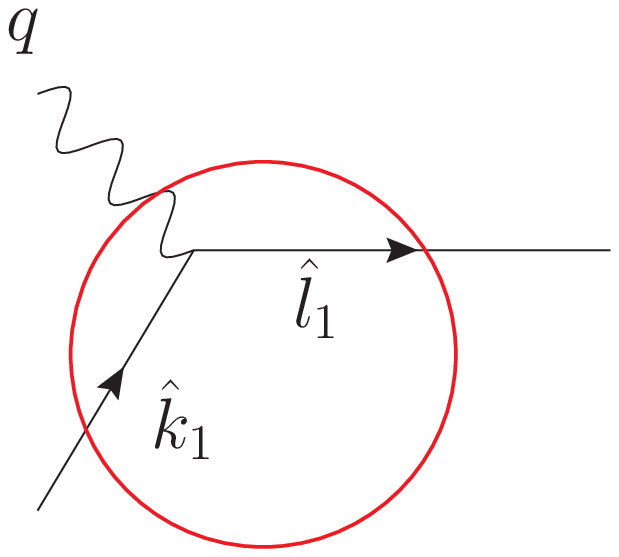}\right|^2,
\end{equation}
is just the $\gamma^{\ast} q \to q$ amplitude squared and averaged over spin.
(Recall that in our convention a factor of $e_j^2$ has already been factored out front in 
Eq.~(\ref{eq:gammasummod}).)
In Eq.~(\ref{eq:grapha_app}) we have used the approximator notation discussed in Sect.~\ref{sec:strategy};
$T_1$ acts on the amplitude in graph (a) by replacing it with the approximate
expression on the right side, appropriate for region $R_1$.  

The trace in the second factor in Eq.~(\ref{eq:grapha_app}) comes from 
expanding $\jetbub$ in a basis of the Dirac algebra
\begin{equation}
\jetbub(l)  =
    \jetbub_S
    + \gamma^{\mu} \jetbub_{\mu} +\sigma^{\mu\nu} 
\jetbub_{\mu\nu} +\gamma_5 \jetbub_{5}+ \gamma^{\mu} \gamma_5\jetbub_{\mu 5} \label{eq:cliffordaa}, \\
\end{equation} 
and recognizing that the dominant term in unpolarized scattering is $\jetbub^-(l_1) \gamma^+$.
This is the term projected out by $\mathcal{P}_T$ and $\mathcal{P}_J$ in Eqs.~(\ref{eq:proj1}).
The minus component of $\jetbub(l_1)$ is determined from the trace
\begin{equation}
\label{eq:jetproj}
\jetbub^-(l_1) = \frac{1}{4} {\rm Tr} \left[ \gamma^- \jetbub(l_1) \right]. 
\end{equation}
In the second line of Eq.~(\ref{eq:grapha_app}), we have used braces to help distinguish 
the basic elements of the factorized expression.  
Once the integral over $l_2$ is included, the last factor in braces is identifiable as an order-$\strong$ contribution 
to the fully unintegrated quark PDF, Eq.~(\ref{quarkpcf}).
The simple partonic vertex Eq.~(\ref{eq:vertex}) has been pulled out in front.
We have extracted an over-all factor of $1/ Q^2$ to give the conventional normalization of the $\gamma^\ast q \rightarrow q$ amplitude.  
The factorized form in Eq.~(\ref{eq:grapha_app}) resulting from the action of $T_1$ on Eq.~(\ref{eq:gammasummod}) is 
illustrated by the graph on the right side of the arrow in Fig.~\ref{fig:grapha_app}. 
To simplify notation, the factored bubbles in Fig.~\ref{fig:grapha_app} implicitly include the traces in Eq.~(\ref{eq:grapha_app}).
Hence, to simplify later expressions we define
\begin{eqnarray}
J(l_1) = {\rm Tr} \left[ \gamma^- \jetbub(l_1) \right] = \picineq{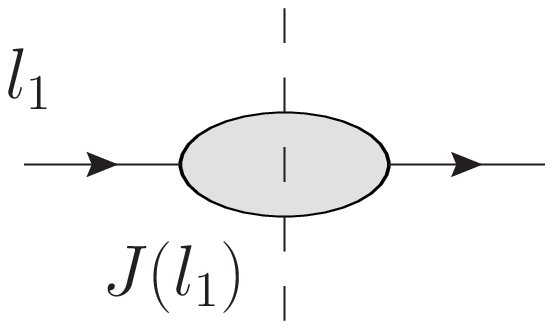} \label{LOjet1} \\
J(l_2) = {\rm Tr} \left[ \gamma^- \jetbub(l_2) \right] = \picineq{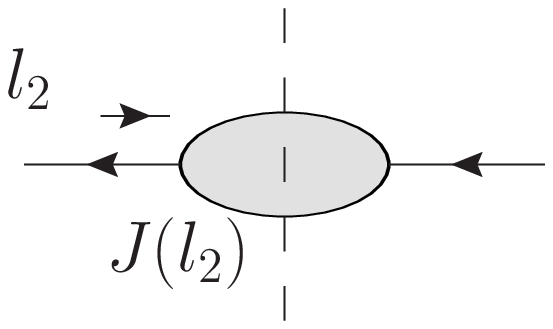}. \label{LOjet2}
\end{eqnarray}

\subsection{Figure~\ref{fig:boxdiagrams}(b)}
For gauge theories, the situation is more complicated than, for example, the six-dimensional scalar-$\phi^3$ theory used in Ref.~\cite{CZ}
because, in addition to graphs like Fig.~\ref{fig:boxdiagrams}(a), we can also have leading contributions from graphs like Fig.~\ref{fig:boxdiagrams}(b) where there is both a target collinear 
quark and a target collinear gluon attaching to the hard part (see the left side of the arrow in Fig.~\ref{fig:graphb_app}).  The extra collinear gluon can be disentangled with the 
help of a Ward identity argument as we now illustrate explicitly.
The unapproximated 
expression is
\begin{equation}
\label{eq:ampunapproxb}
\left( \left| M_{b}(l_1,l_2,k) \right|^2 \right)^{\mu \nu} = \strong T_R {\rm Tr} \left[ \gamma^\nu 
\jetbub(l_1) \gamma^\kappa \left( \frac{1}{\slashed{l}_1 - \slashed{k} - m} \right) \gamma^{\mu} \jetbub(l_2) 
\gamma^\rho \left( \frac{1}{\slashed{k}_1 - m} \right) \right] \Phi_{\kappa \rho}(k,P).
\end{equation}
Following step 2 of Sect.~\ref{sec:strategy}, we draw solid red circles around the hard 
parts as shown in Fig.~\ref{fig:graphb_app}.
\begin{figure*}
\centering
    \includegraphics[scale=0.6]{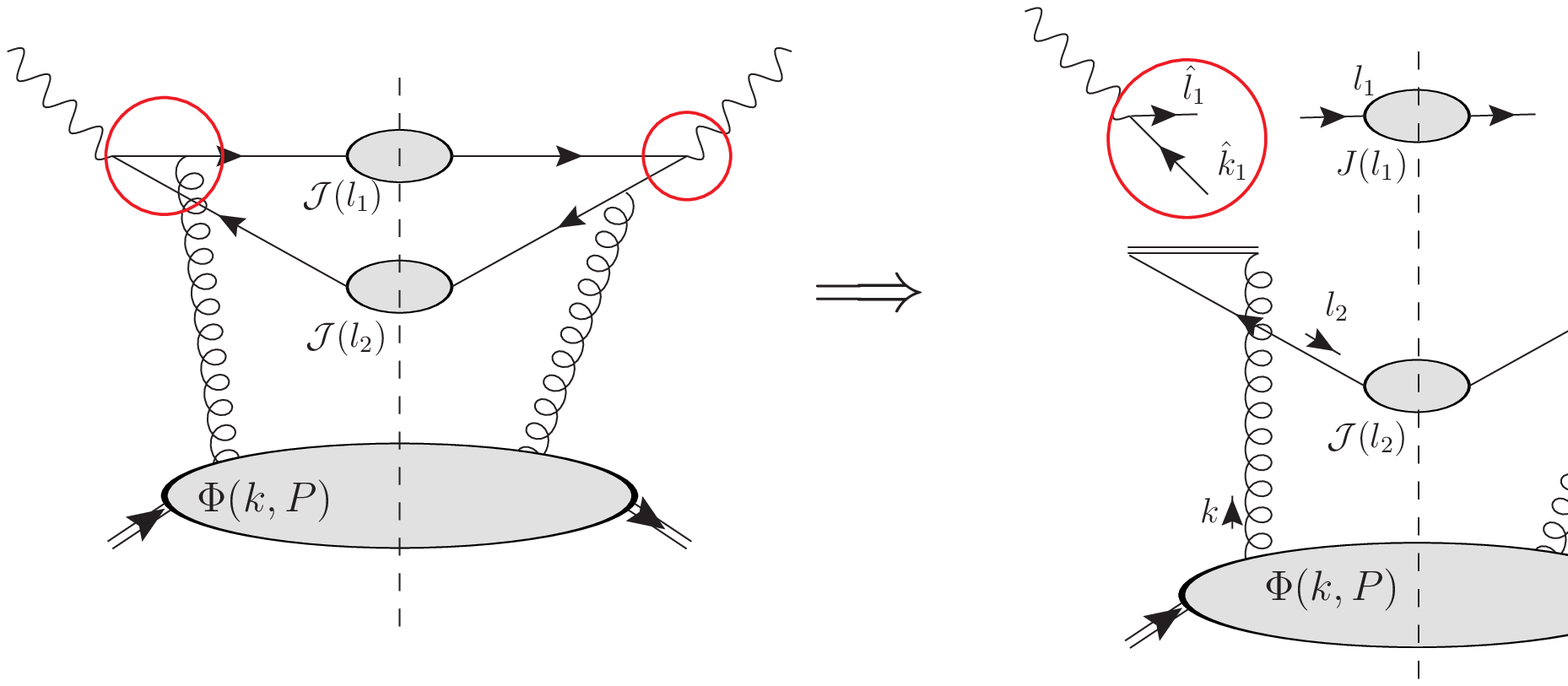}
\caption{Figure~\ref{fig:boxdiagrams}(b) with the red circles showing the application of the target collinear approximation.  The graph on the right hand side 
of the arrow shows the separation into the factors of Eq.~(\ref{eq:pmhardfactb}).  The gluon is shown attaching to an eikonal line represented by double lines.}
\label{fig:graphb_app}
\end{figure*}
On the right side of the cut, we make the same approximation as in graph (a).
However, on the left side, we must first dot $\approxb{k}$ into the hard bubble and multiply 
by $n_s^{\kappa} / (n_s \cdot k - i\epsilon)$, as dictated by the prescription in section XB of Ref.~\cite{CRS}.  Inside the hard circles, the prescription of Ref.~\cite{CRS} is to replace momentum 
variables by
\begin{align}
\label{eq:approxb2}
\approxb{l}_{1} = (0,q^-,\3{0}_\trans ), \qquad
\approxb{k} =& \left( \frac{-q^+ k^+}{k^+ - l_2^+},0,\3{0}_\trans  \right), 
\qquad -\approxb{l}_2 = \left( \frac{q^+ l_2^+}{k^+ - l_2^+},0,\3{0}_\trans  \right).
\end{align}
Note that $\approxb{k}_1 = \approxb{k} - \approxb{l}_2$, which is consistent with the requirement that 
approximate momentum is conserved.
The solid red circles therefore tell us to make the replacements
\begin{equation}
\label{eq:repb}
\gamma^{\kappa} \left( \frac{1}{\slashed{l}_1 - \slashed{k} - m} \right) \gamma^{\mu} \to
\frac{n_s^{\kappa}}{n_s \cdot k - i\epsilon} \mathcal{P}_J 
\slashed{\approxb{k}} \left( \frac{1}{\slashed{\approxb{l}}_1 - \slashed{\approxb{k}}} \right) \gamma^{\mu} \mathcal{P}_J, 
\qquad \gamma^{\nu} \to \mathcal{P}_T \gamma^\nu \mathcal{P}_T.
\end{equation}
Note that it is necessary to use the precise definition of the replacements in Eqs.~(\ref{eq:approxb2},\ref{eq:repb}) 
to remain consistent with the factorization formalism of Ref.~\cite{CRS}.
The contribution from graph (b) to the complete amplitude thus becomes
\begin{equation}
\left( \left| M_{b} \right|^2 \right)^{\mu \nu} \to
\frac{\strong T_R}{(2 Q^2)^2} {\rm Tr} \left[ \gamma^\nu 
\slashed{\approxb{l}}_1 \slashed{\approxb{k}} \left( \frac{1}{\slashed{\approxb{l} }_1 
- \slashed{\approxb{k}}} \right) \gamma^{\mu} \slashed{\approxb{k}}_1 \right] {\rm Tr} \left[ 
\slashed{\approxb{k}}_1 \jetbub(l_1) \right] 
\frac{n_s^{\kappa} }{n_s \cdot k - i\epsilon} 
{\rm Tr} \left[ \slashed{\approxb{l}}_1  \jetbub(l_2) \gamma^\rho \left( \frac{1}{\slashed{k}_1 - m} \right) \right] 
\Phi_{\kappa \rho}(k,P).
\label{eq:graphb}
\end{equation}
A Ward identity argument now allows the collinear gluon to be factored away from the hard part;
substituting $\slashed{\approxb{k}} = -(\slashed{\approxb{l}}_1 - \slashed{\approxb{k}}) 
+ \slashed{\approxb{l}}_1$ into Eq.~(\ref{eq:graphb}) leads to
\begin{align}
\label{eq:pmhardfactb}
T_1 \left( \left| M_{b} \right|^2 \right)^{\mu \nu} =
 -\frac{\strong T_R}{(2 Q^2)^2} {\rm Tr} \left[ \gamma^\nu 
\slashed{\approxb{l}}_1 \gamma^{\mu} \slashed{\approxb{k}}_1 \right] {\rm Tr} \left[ 
\slashed{\approxb{k}}_1 \jetbub(l_1) \right] 
\frac{n_s^{\kappa} }{n_s \cdot k - i\epsilon} 
{\rm Tr} \left[ \slashed{\approxb{l}}_1  \jetbub(l_2) \gamma^\rho \left( \frac{1}{\slashed{k}_1 - m} \right) \right] 
\Phi_{\kappa \rho}(k,P) \nonumber \\
=  \frac{1}{Q^2}  
\left( \overline{\left| A^{\gamma^{\ast} q \rightarrow q}
(\approxb{l}_1,\approxb{k}_1) \right|}^2 \right)^{\mu \nu}
\left\{ {\rm Tr} \left[ \gamma^- \jetbub(l_1) \right] \right\}
\left\{ - \strong T_R \frac{n_s^{\kappa} }{n_s \cdot k - i\epsilon} 
{\rm Tr} \left[ \frac{\gamma^+}{4}  \jetbub(l_2) \gamma^\rho \left( \frac{1}{\slashed{k}_1 - m} \right) \right] 
\Phi_{\kappa \rho}(k,P) \right\}.
\end{align}
\subsection{Figures~\ref{fig:boxdiagrams}(c) and~\ref{fig:boxdiagrams}(d)}

The remaining graphs in Fig.~\ref{fig:graphb_app} follow exactly similar steps as
graphs (a) and (b).  
For graph (c) we have
\begin{multline}
T_1 \left( \left| M_{c} \right|^2 \right)^{\mu \nu} = \\
\frac{1}{Q^2} \left( \overline{\left| A^{\gamma^{\ast} q \rightarrow q}
(\approxb{l}_1,\approxb{k}_1) \right|}^2 \right)^{\mu \nu}
\left\{ {\rm Tr} \left[ \gamma^- \jetbub(l_1) \right] \right\}
\left\{ - \strong T_R \frac{n_s^{\rho} }{n_s \cdot k + i\epsilon} 
{\rm Tr} \left[ \frac{\gamma^+}{4}  \left( \frac{1}{\slashed{k}_1 - m} \right)  \gamma^{\kappa} \jetbub(l_2) \right] 
\Phi_{\kappa \rho}(k,P) \right\},
\end{multline}
and for graph (d)
\begin{multline}
T_1 \left( \left| M_{d} \right|^2 \right)^{\mu \nu} = \\
\frac{1}{Q^2} \left( \overline{\left| A^{\gamma^{\ast} q \rightarrow q}
(\approxb{l}_1,\approxb{k}_1) \right|}^2 \right)^{\mu \nu}
\left\{ {\rm Tr} \left[ \gamma^- \jetbub(l_1) \right] \right\}  
\left\{ \strong T_R \frac{ n_s^{\kappa} }{n_s \cdot k - i\epsilon} \frac{ n_s^{\rho} }{n_s \cdot k + i\epsilon} 
{\rm Tr} \left[ \frac{\gamma^+}{4} \jetbub(l_2) \right]
\Phi_{\kappa \rho}(k,P) \right\}.
\end{multline}

\subsection{The sum of graphs (a)-(d)}

Adding graphs (a-d) results in a factorized structure which we can express in terms of diagrams as
\begin{multline}
\label{eq:factored1}
T_1 \left( \left| M(l_1,l_2,k) \right|^2 \right)^{\mu \nu} 
= 
\frac{1}{Q^2} \left| \picineq{vertex} \right|^2 \times \left( \picineq{jetbubLO} \right) \times \\ \times \left( \picineq{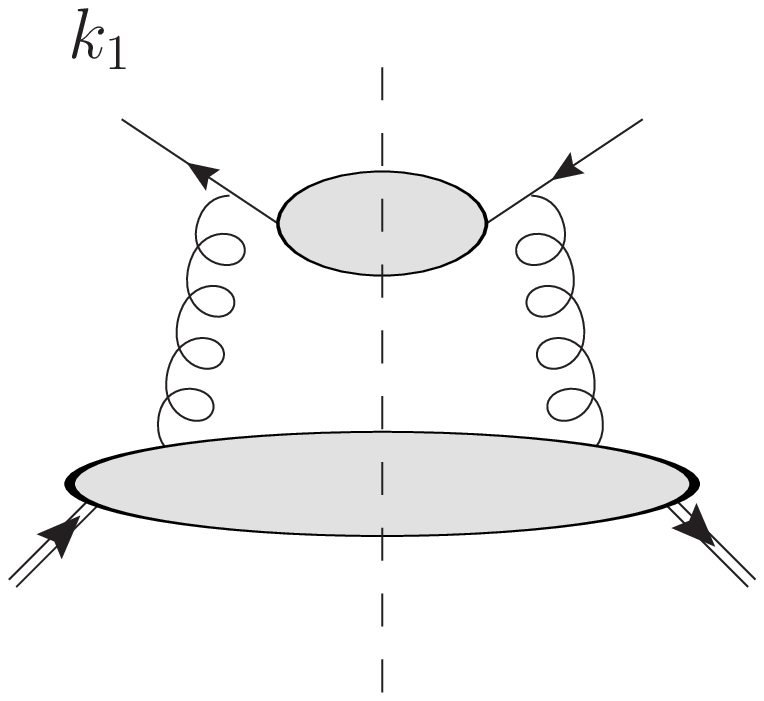} + \picineq{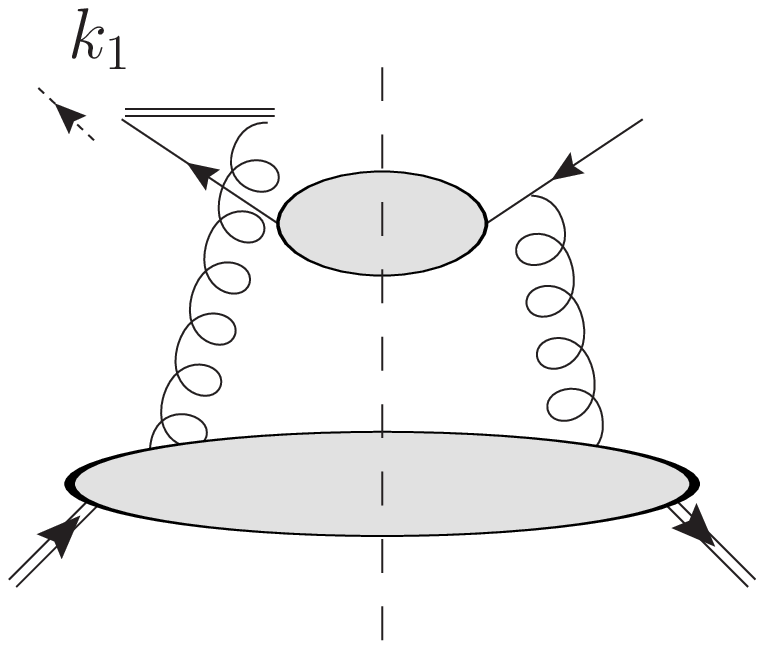} + \picineq{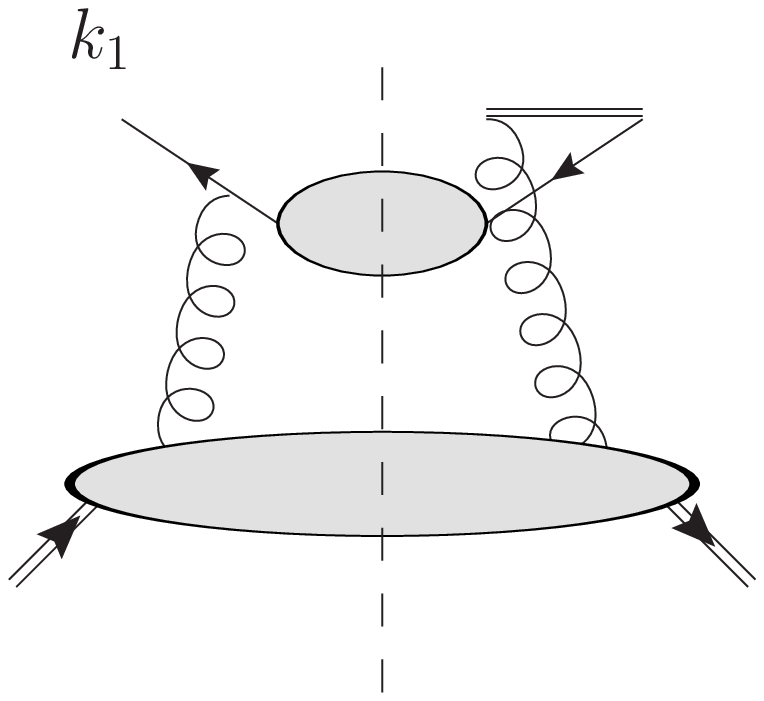} + \picineq{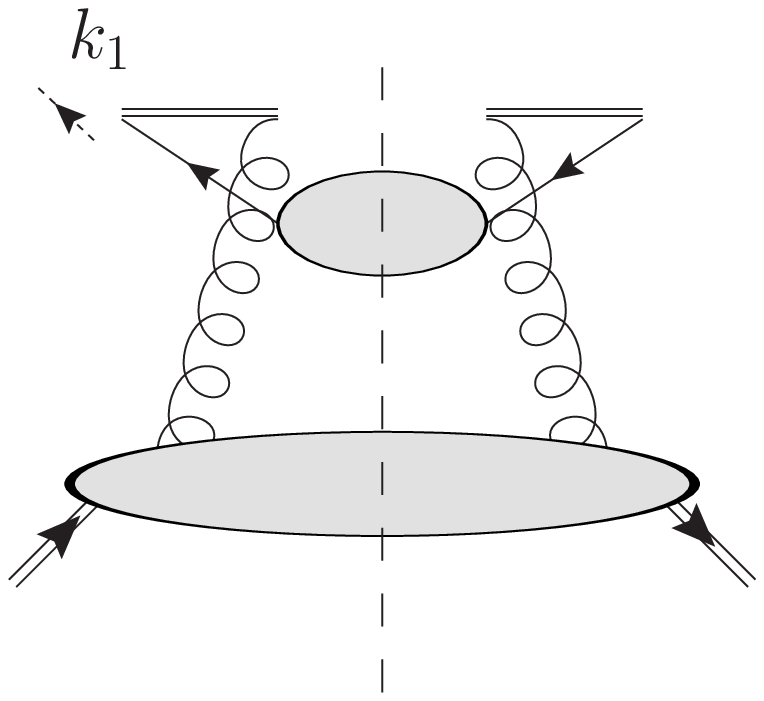} \right).
\end{multline}
The graphs in the first two factors in Eq.~(\ref{eq:factored1}) are the same as in Eqs.~(\ref{eq:vertex},\ref{eq:jetproj}).
They do not depend 
on $l_2$.  Hence, we can push the $l_2$ integral in Eq.~(\ref{eq:gammasummod}) through to integrate
over the last factor in parentheses and obtain the single real anti-quark contribution to the 
fully unintegrated quark PDF,
\begin{multline}
\label{eq:quarkpdf}
\pdf^{(1\bar{q})}_{q/p}(k_1,P) = 
\int \frac{d^4 l_2}{(2 \pi)^4} \; \strong T_R \left\{ {\rm Tr} \left[ \frac{\gamma^+}{4} \left( \frac{1}{\slashed{k}_1 - m} \right) \gamma^{\kappa} \jetbub(l_2) 
\gamma^\rho \left( \frac{1}{\slashed{k}_1 - m} \right) \right] \right. - \\ - 
\frac{n_s^{\kappa} }{n_s \cdot k} 
{\rm Tr} \left[ \frac{\gamma^+}{4}  \jetbub(l_2) \gamma^\rho \left( \frac{1}{\slashed{k}_1 - m} \right) \right]  -
\frac{n_s^{\rho}}{n_s \cdot k} 
{\rm Tr} \left[ \frac{\gamma^+}{4}  \left( \frac{1}{\slashed{k}_1 - m} \right)  \gamma^{\kappa} \jetbub(l_2) \right] + \\ +
\left. \frac{n_s^{\kappa} }{n_s \cdot k} \frac{n_s^{\rho} }{n_s \cdot k} 
{\rm Tr} \left[ \frac{\gamma^+}{4} \jetbub(l_2) \right] \right\} \Phi_{\kappa \rho}(k,P).
\end{multline}
The superscript $(1\bar{q})$ symbolizes that these diagrams represent the contribution 
to the fully unintegrated quark PDF due to the hadronization of a single real anti-quark in the final state.
Note the double lines in the last factor of Eq.~(\ref{eq:factored1}) denoting the 
eikonal propagators of Eq.~(\ref{eq:quarkpdf}).
Equation~(\ref{eq:quarkpdf}) is precisely what arises from an expansion of the operator definition in Eq.~(\ref{quarkpcf}) for the fully unintegrated 
quark PDF.  (See Ref.~\cite{CSSreview,CRS} for more details on writing down Feynman rules for the quark PDF.)
In Eq.~(\ref{eq:quarkpdf}) and from here on out, we drop the explicit $i\epsilon$ terms in the eikonal propagators to simplify the expressions.
Given models (or fits to data) for $\jetbub$ and $\Phi$, we can explicitly 
calculate the contribution from Figs.~\ref{fig:boxdiagrams} to the fully unintegrated quark PDF in perturbation theory using Eq.~(\ref{eq:quarkpdf}).  Of course, 
perturbative methods are only valid for a large enough $k_1^2$.
Near the core of region $R_1$ ($k_1$ with almost exactly vanishing minus and transverse components), the fully unintegrated quark PDF should be modeled or fit 
to data.  Additionally, there is freedom in how one chooses the precise value of $y_s$ so long as $y_s \approx 0$.


\subsection{Region $\bar{R}_1$: Target anti-quark}
If we consider the case that the target collinear parton is an anti-quark, then instead of 
Eq.~(\ref{eq:regionR1}) we use kinematic approximations appropriate to region $\bar{R}_1$:
\begin{equation}
k_1^\prime \sim \left(Q,\Lambda^2/Q,\Lambda \right),
\end{equation}
where (see graph (d) of Fig.~\ref{fig:boxdiagrams}) recall that
\begin{equation}
k_1^\prime = l_1 - k.
\end{equation}
We define the approximate variables analogously to Eq.~(\ref{eq:core})
for the target quark case.  This time it is the anti-quark which gains a 
large minus component after being struck by the virtual photon.  So, using an inverted ``hat'' to symbolize approximate 
variables for region $\bar{R}_1$, we have
\begin{equation}
\label{eq:corebar}
\check{k}_1^\prime = (-q^+,0,\3{0}_t), \qquad \check{l}_2 = (0,q^-,\3{0}_t ).
\end{equation}
The remaining steps are exactly analogous to those of the treatment 
of the quark target in the previous subsection.

\subsection{Summary: Factorization formula for $\gamma^\ast p \to {\rm 1 \; \; Jet} + X$ scattering.}

We can summarize the results of this section by writing the explicit factorization formula 
for the contribution to the hadronic structure tensor from $\gamma^\ast p \to J_{q(\bar{q})} + X$ scattering.
(The subscript $q (\bar{q})$ on $J$ indicates that the jet arises from a knocked-out quark (anti-quark).)
Using Eq.~(\ref{eq:factored1}) inside Eq.~(\ref{eq:gammasummod}) and using the $l_1$ integral 
to evaluate the $\delta$-function, we have
\begin{equation}
\label{eq:T1}
\hadtensor^{\mu \nu}_{\gamma^\ast p \to J_q + X}(P,q) = \frac{e_j^2}{4 \pi} \int \frac{d^4 k_1}{(2 \pi)^4} \frac{1}{Q^2}  \left( \overline{\left| A^{\gamma^{\ast} q \rightarrow q}
(\approxb{l}_1,\approxb{k}_1) \right|}^2 \right)^{\mu \nu} \, J(l_1) \, \pdf_{q/p}(k_1,P).
\end{equation}
We have changed variables so that the integration in Eq.~(\ref{eq:T1}) is over struck quark momentum, $k_1$.
Furthermore, we have dropped the $(1\bar{q})$ subscript on the fully unintegrated PDF in Eq.~(\ref{eq:T1}) because, in general,
we need to include all graphs that contribute to the quark PDF in addition to those shown in Eq.~(\ref{eq:quarkpdf}).
Note that the non-perturbative functions, $J(l_1)$ and $\pdf_{q/p}(k_1,P)$ are evaluated using \emph{exact} momentum.
The procedure for evaluating Eq.~(\ref{eq:T1}) can be summarized as follows:
\begin{itemize}
\item Obtain parameterizations of $J(l_1)$ and $\pdf_{q/p}(k_1,P)$ from models or fits.
\item Evaluate the LO hard scattering coefficient, $A^{\gamma^\ast q \to q}$, using Eqs.~(\ref{eq:core}) to 
obtain $\approxb{l}_1$ and $\approxb{k}_1$. 
\end{itemize}

The analogous expression for $\bar{R}_1$ is
\begin{equation}
\label{eq:T1bar}
\bar{\hadtensor}^{\mu \nu}_{\gamma^\ast p \to J_{\bar{q}} + X}(P,q) = \frac{e_j^2}{4 \pi} \int \frac{d^4 k_1^\prime}{(2 \pi)^4} \frac{1}{Q^2} \left( \overline{\left| A^{\gamma^{\ast} \bar{q} \rightarrow \bar{q}}
(\check{l}_2,\check{k}^\prime_1) \right|}^2 \right)^{\mu \nu} \, J(l_2) \, \bar{\pdf}_{\bar{q}/p}(k_1^\prime,P).
\end{equation}
Given a particular model or parameterization of the fully unintegrated jet factor, and the fully unintegrated quark PDF, 
Eqs.~(\ref{eq:T1},\ref{eq:T1bar}) allow a direct calculation of the first two terms in Eq.~(\ref{eq:wideangle}).
Together, they give the complete contribution from the graphs in Fig.~\ref{fig:boxdiagrams} to LO 
hard scattering.
Away from the core of regions $R_1$, one can also estimate $F_{q/p}$ by 
directly calculating Eq.~(\ref{eq:quarkpdf}), given knowledge of $\pdfbub(k,P)$.
For low-angle scattering we should also include the order-$\strong$ virtual correction to the hard vertex.  This was 
already worked out in Ref.~\cite{CRS} so we do not repeat the steps here.

\end{widetext}
\section{Region $R_2$: Large Transverse Momentum}
\label{sec:hardregion}
\begin{figure*}
\centering
  \begin{tabular}{c@{\hspace*{5mm}}c}
    \includegraphics[scale=0.4]{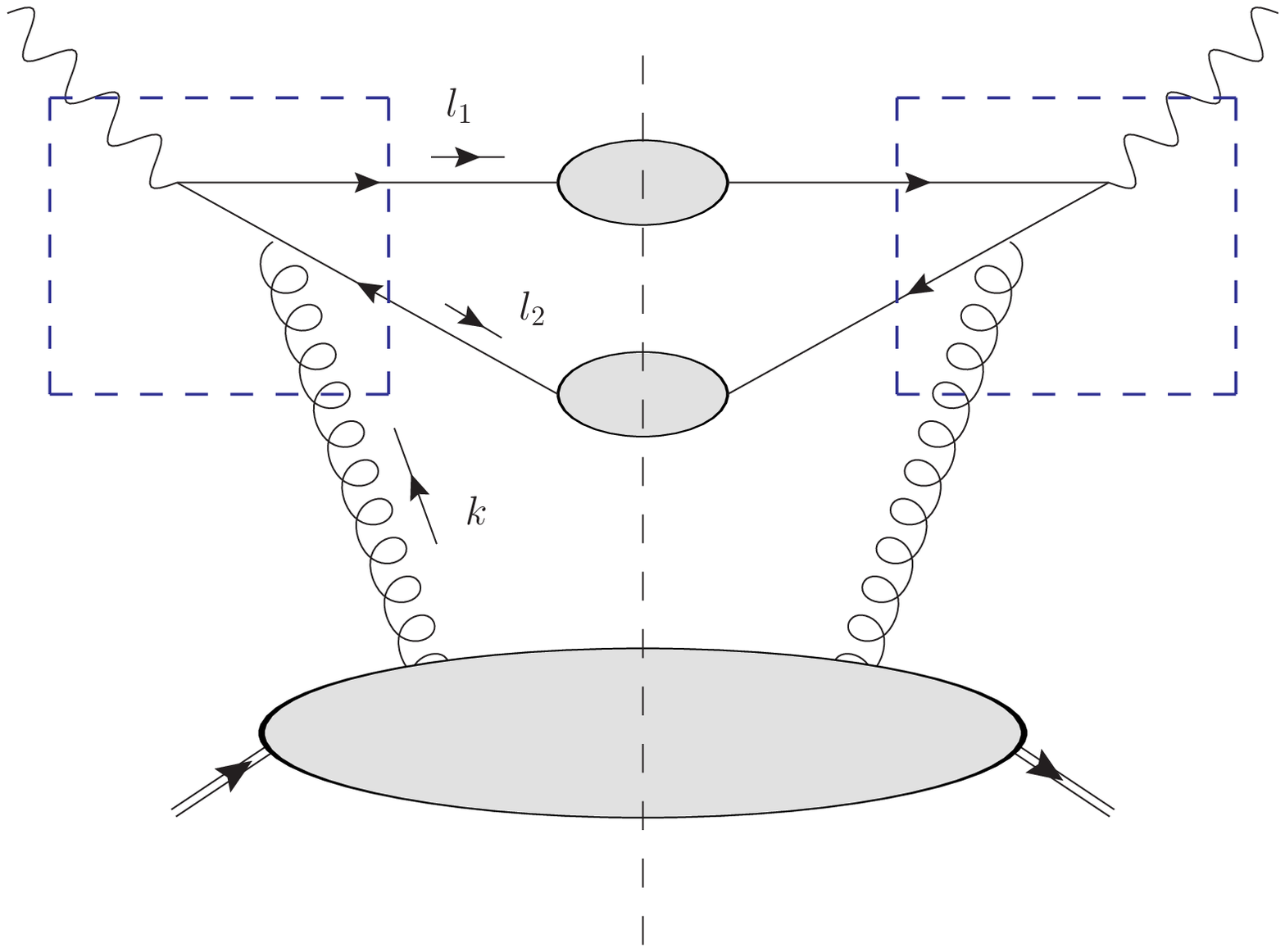}
    &
    \includegraphics[scale=0.4]{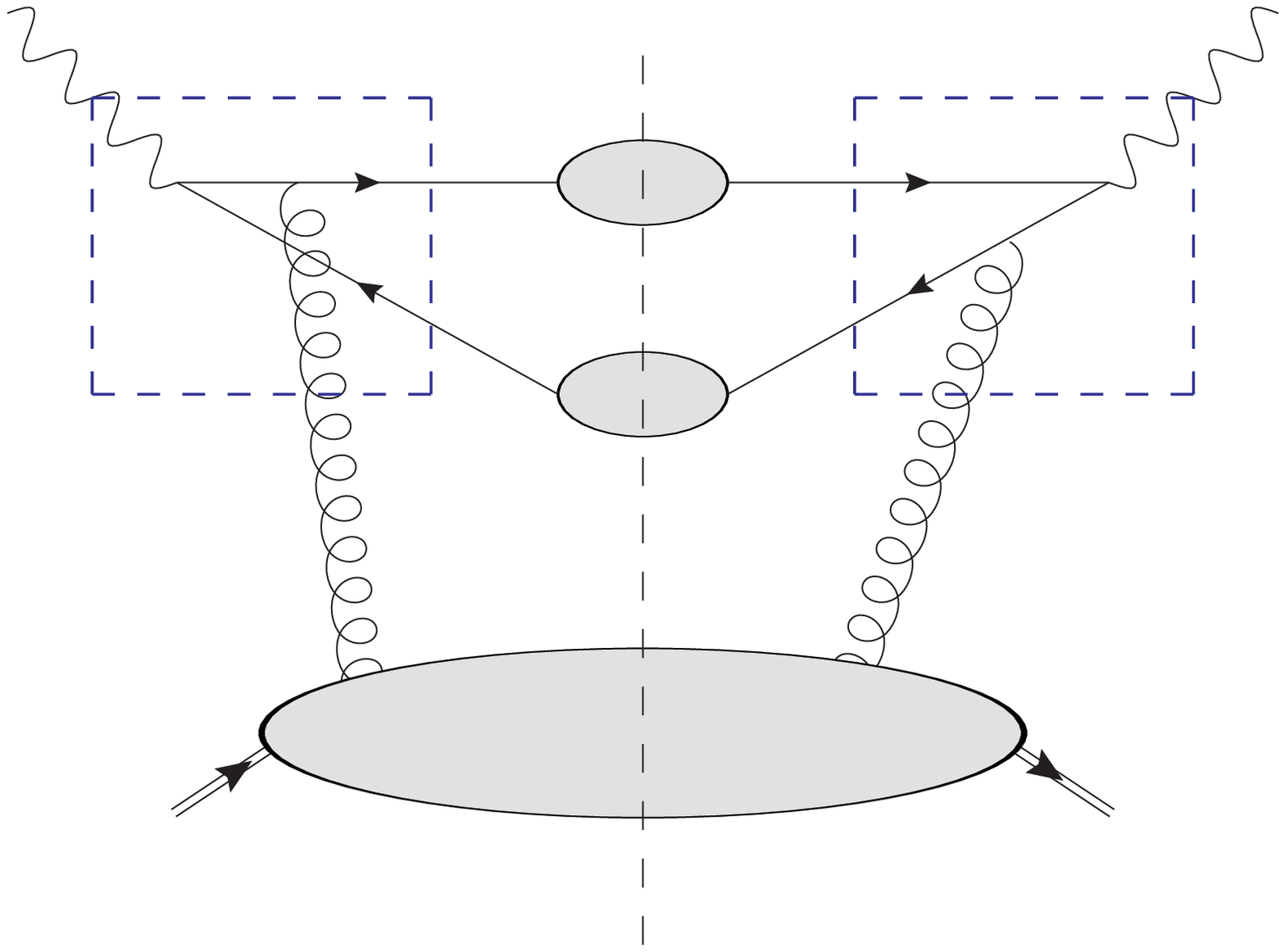}
  \\
    $(a)$ & $(b)$
  \end{tabular}
\caption{Dashed blue boxes symbolizing the application of the approximation appropriate
for region $R_2$ to Figs.~\ref{fig:boxdiagrams}(a) and (b).  Analogous graphs are needed for Figs.~\ref{fig:boxdiagrams}(c) and (d), though we do not show them here explicitly.}
\label{fig:blueboxes}
\end{figure*}

The next step is to consider region $R_2$, where the virtuality of 
$k_1$ is large and Eq.~(\ref{eq:core}) is a poor approximation.
However, $k$ still has nearly target-collinear momentum given 
by Eq.~(\ref{eq:ksize}), and $l_1$ and $l_2$ are highly boosted 
relative to one another.  Therefore, we should regard the $\gamma^{\ast} g \rightarrow q \bar{q}$
subgraph as the basic partonic subprocess;
after making approximations appropriate for $R_2$, 
it should be possible to factor the elementary on-shell
amplitude for $\gamma^{\ast} g \rightarrow q \bar{q}$ scattering.
Therefore, the approximated graph in region $R_2$ will
be symbolized diagrammatically by a dashed blue rectangular box 
enclosing the $\gamma^{\ast} g \rightarrow q \bar{q}$ subgraph 
as shown in Fig.~\ref{fig:blueboxes}.  The box should be read as an instruction to replace
all momenta inside it by the appropriate approximated momenta for $R_2$ (to be discussed below).  Furthermore, at 
each intersection of a parton line with the dashed-blue box,
there needs to be an appropriate projection onto 
Lorentz/Dirac components.
The purpose of this section is to establish exactly what these approximations 
must be such that the resulting expression is consistent with the factorization formalism that was set up
in Refs.~\cite{CZ,CRS}.

Returning now to Eq.~(\ref{eq:gammasummod}), we decompose the square modulus of the amplitude
into the contraction of subgraphs
\begin{widetext}
\begin{equation}
\label{eq:decomp1}
\left( \left| M(l_1,l_2,k) \right|^2 \right)^{\mu \nu} = {\rm Tr} 
\left[ \jetbub(l_1) \hardbub^{\mu, \kappa}_{L}(l_1,l_2,k) 
\jetbub(l_2) \hardbub^{\nu, \kappa^{\prime}}_{R}(l_1,l_2,k) \right] \pdfbub_{\kappa,\kappa^{\prime}}(k,P). 
\end{equation}
\end{widetext}
The $\hardbub_{L/R}$ are now the hard subgraphs for region $R_2$, i.e., the 
subgraphs enclosed by dashed blue boxes of Fig.~\ref{fig:blueboxes}.
To simplify notation, Dirac indices will be left implicit.

\subsection{Approximations on kinematic variables}
\label{sec:kinematics}

Inside the dashed blue boxes, we replace the exact parton momentum variables
with approximate momentum
variables appropriate for the production of two wide-angle
jets with large transverse momentum.
We use the approximation scheme proposed in~\cite{CZ} 
applied here to the special case of two out-going jets.
The choice of a good approximation scheme is probably not unique, though
once one is chosen it should be maintained throughout the rest of the calculation.
The approximate momentum variables 
for region $R_2$ will be denoted by a tilde placed over the corresponding unapproximated variable.

Enforcing partonic values for the approximate
momenta amounts to requiring
\begin{align}
\label{eq:part}
\approxa{k} =& (\approxa{k}^+,0,\3{0}_{\trans} ), \qquad \
\approxa{l}_1^2 = 0, \qquad \approxa{l}_2^2 = 0.
\end{align}
The true values of $\3{k}_{\trans}$, $k^-$, $l_1^2 = M_1^2$ and $l_2^2 = M_2^2$ therefore
parameterize the deviation of the approximate kinematics from exact kinematics.  
In addition, we require that both the exact and approximate momenta 
obey total four-momentum conservation:
\begin{align}
\label{eq:cons}
\approxa{k} + q = \approxa{l}_1 + \approxa{l}_2, \\
\label{eq:consb}
k + q = l_1 + l_2.
\end{align}
A choice of a mapping between exact and
approximate momenta that satisfies these constraints is~\cite{CZ}
\begin{align}
\label{eq:approxmoma}
\approxa{k} =& \left( -q^+ + \frac{\approxa{l}_{1,\trans}^{2}}{2 \approxa{l}^-_1} + 
\frac{\approxa{l}_{2,\trans}^{2}}{2 \approxa{l}^-_2},0,\3{0}_{\trans} \right) \\
\label{eq:approxmomab}
\approxa{l}_{1} =& \left( \frac{\approxa{l}_{\trans,1}^2}{2 \approxa{l}^-_1}, 
\tilde{l}^-_1 ,\3{\approxa{l}}_{1,\trans} \right) \\ 
\label{eq:approxmomac}
\approxa{l}_{2} =& \left( \frac{\approxa{l}_{\trans,2}^2}{2 \approxa{l}^-_2}, 
\tilde{l}^-_2 ,\3{\approxa{l}}_{2,\trans} \right),
\end{align}
where
\begin{align}
\3{\approxa{l}}_{1,\trans} =& \3{l}_{1,\trans} -\3{k}_{\trans}/2, \qquad  
\3{\approxa{l}}_{2,\trans} = \3{l}_{2,\trans} - \3{k}_{\trans}/2; \\ 
\label{eq:approxe}
\approxa{l}_{1}^- =& l_{1}^- - k^{-}/2, \qquad \approxa{l}_{2}^- = l_{2}^- - k^{-}/2.  
\end{align}
In the limit of small $\3{k}_{\trans}$ and small $k^-$, we have $\3{\approxa{l}}_{j,\trans} \approx \3{l}_{j,\trans}$ and $\approxa{l}_{j}^- \approx l_{j}^-$.
Again, the choice of approximation scheme in Eqs.~(\ref{eq:approxmoma}-\ref{eq:approxe}) is 
not uniquely determined from Eqs.~(\ref{eq:part}-\ref{eq:consb}).
Other equally good approximation schemes may have advantages, but the one used here and in Ref.~\cite{CZ} is
sufficient for our purposes.
With the constraints of Eq.~(\ref{eq:cons},\ref{eq:consb}) we also have 
$\3{\approxa{l}}_{1,\trans} = - \3{\approxa{l}}_{2,\trans}$.
Given the exact values of initial and final state momentum, we can write 
the inverse transformations
\begin{align*}
  l_{j, \trans} =& \approxa{l}_{j, \trans} + \frac{k_t}{2}, \qquad l_j^- = 
\approxa{l}^-_{j} + \frac{k^-}{2}; \\
l_j^+ =& \frac{l_{j, \trans}^2 + M_j^2}{2 l_j^-}, \qquad k^+ = \sum_j l_j^+ - q^+.
\end{align*}
Here $j = 1,2$ label the out-going quark and anti-quark jets.
We therefore have a one-to-one mapping between exact and approximate momentum variables for region $R_2$.
It will also be useful to define approximate versions of the partonic Mandelstam variables:
\begin{equation}
\label{eq:approxman}
\approxa{s} = (q + \approxa{k})^2, \qquad \approxa{t} = (\approxa{l}_2 - \approxa{k})^2, \qquad \approxa{u} = (\approxa{l}_1 - \approxa{k})^2.
\end{equation}

Our main goal is to apply a sequence of approximations that allow
the unpolarized, on-shell square modulus of the partonic amplitude to be factored
out of the complete sum of graphs in Fig.~\ref{fig:boxdiagrams}.
Part of the approximation is, of course, to replace $l_1$,$l_2$, and $k$
by their approximate values $\approxa{l}_1$, $\approxa{l}_2$, and $\approxa{k}$,
defined above.  That is, we will ultimately need to make the replacement
\begin{equation}
\label{eq:gluapp1}
\hardbub^{\mu , \kappa}_{L,R}(l_1,l_2,k) \to
\hardbub^{\mu , \kappa}_{L,R}(\approxa{l}_1,\approxa{l}_2,\approxa{k}).
\end{equation}
However, in a gauge
theory we also need to project out the relevant Lorentz/Dirac 
components of the external lines, in a way that is appropriate for unpolarized scattering with massless on-shell partons.  
The steps for these projections will be discussed in the next few subsections.

\subsection{Target gluon polarization}

\begin{figure*}
\centering
    \includegraphics[scale=0.4]{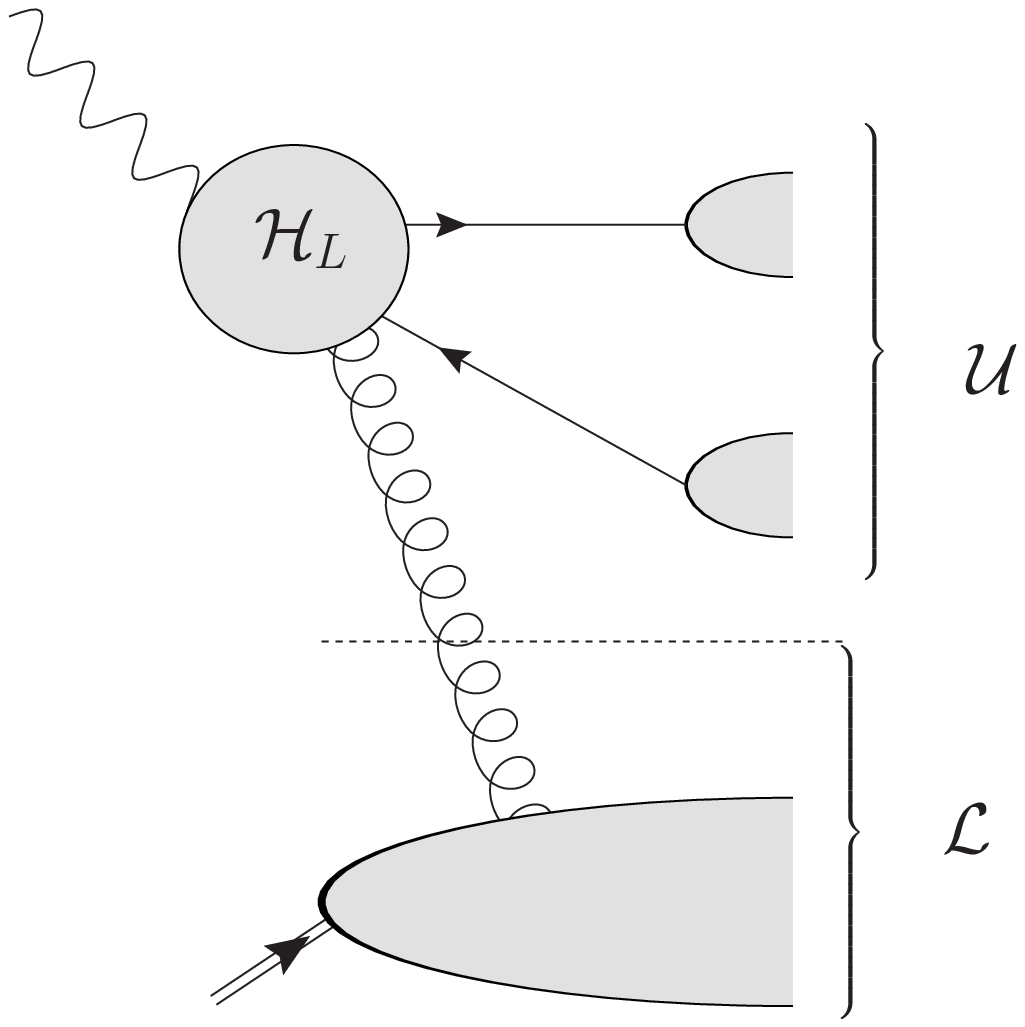}
\caption{Basic amplitude separated into an upper part $\upperbub$ and a lower part $\pdfamp$, see Eq.~(\ref{eq:amp})}
\label{fig:amplitude}
\end{figure*}

To factorize the fully unintegrated gluon PDF, we will first need to 
project out the relevant polarizations of the target gluon.
It will be convenient 
to work at the level of the amplitude, decomposed
into upper and lower blocks as shown in Fig.~\ref{fig:amplitude}.
The sum of graphs contributing above the dotted 
line will be called $\upperbub$ and the sum 
of graphs below the dotted line will be called $\pdfamp$.
The steps for projecting gluon polarizations will follow closely 
those in Ref.~\cite{CR} for the standard integrated PDF.
Decomposing the amplitude in this way, we have
\begin{equation}
\label{eq:amp}
M^{\mu}(l_1,l_2,k) =  \upperbub^{\mu}_{\kappa}(l_1,l_2,k) \, g^{\kappa \kappa^\prime} \, \pdfamp_{\kappa^\prime}(k,P),
\end{equation}
Here we have explicitly extracted the numerator, $g^{\mu \nu}$, of the gluon propagator.
We will neglect the electromagnetic indices 
and the momentum arguments in this expression when convenient.

The projections on target gluon polarizations
should allow the lower part of the 
graph 
to be separated into a factor that is 
consistent with a definition for the fully unintegrated gluon PDF.
To this end, we note that, in the center of mass frame, the target bubble is highly boosted in the plus direction so 
the components of the lower bubble have relative sizes given by, 
\begin{align}
\label{eq:lowersize}
\frac{\pdfamp_\trans}{\pdfamp^+} = \mathcal{O}\left( \frac{\Lambda}{Q} \right), 
\qquad \frac{\pdfamp^-}{\pdfamp_\trans} = \mathcal{O}\left( \frac{\Lambda}{Q} \right). 
\end{align}
Since none of the lines in the
upper part of the graph are collinear, all of the components 
of $\upperbub$ are comparable:
\begin{equation}
\label{eq:uppersize}
\upperbub^- \sim \upperbub^+ \sim \upperbub_\trans.
\end{equation}
Given Eqs.~(\ref{eq:lowersize},\ref{eq:uppersize})
it is tempting to keep only the term 
$\upperbub^-\pdfamp^+$ in~(\ref{eq:amp}). 
On a graph-by-graph basis, however, this
leads to the appearance of super-leading contributions 
(i.e., contributions to  
structure functions that vary as $Q^2$ -- see, e.g.,  Ref.~\cite{LS}).  
These super-leading contributions
correspond to scalar and longitudinal polarizations
whose total contribution 
cancels exactly in the sum over all graphs, at least for the case of 
a single target gluon.  
Therefore, they should be systematically removed graph-by-graph.  
We do this, following the 
Grammer and Yennie treatment treatment in QED~\cite{Grammer:1973db},
by first writing the gluon propagator as a sum of two terms which we call ``$K$-terms'' and ``$G$-terms,''
\begin{equation}
\label{eq:gramyen}
g^{\mu \nu} = K^{\mu \nu} + G^{\mu \nu}
\end{equation}
where
\begin{eqnarray}
\label{eq:Kterm}
K^{\mu \nu} & = & \frac{k^{\mu} n_s^{\nu}}{k \cdot n_s} \\
\label{eq:Gterm}
G^{\mu \nu} & = & g^{\mu \nu} - \frac{k^{\mu} n_s^{\nu}}{k \cdot n_s}.
\end{eqnarray}
We use $n_s$ in these definitions because this suggests 
a fully unintegrated gluon PDF density that is closely analogous to the quark 
density defined in Ref.~\cite{CRS}.
The $K$-term in Eq.~(\ref{eq:gramyen}) vanishes exactly in the sum over all contributions to the upper bubble in Eq.~(\ref{eq:amp}), so we drop it 
in our calculations.  
Therefore, we replace Eq.~(\ref{eq:amp}) with
\begin{multline}
\label{eq:amp2}
M(l_1,l_2,k) =  \upperbub_{\kappa}(l_1,l_2,k) \, G^{\kappa \kappa^\prime} \, \pdfamp_{\kappa^\prime}(k,P) = \\ \upperbub^{\kappa}(l_1,l_2,k) \, \tilde{\pdfamp}_{\kappa}(k,P).
\end{multline}
In the last equality, we define $\tilde{\pdfamp}$ as a short-hand for the 
contraction of $\pdfamp_{\kappa^\prime}$ with a $G$-term.
Only the transverse components are unsuppressed in Eq.~(\ref{eq:amp2}).
Recalling Eqs.~(\ref{eq:ksize},\ref{eq:lowersize})
we can verify this directly by checking the relative sizes of each combination of indices  
\begin{align}
\upperbub^+\tilde{\pdfamp}^- \sim & \mathcal{O} \left( \frac{\Lambda}{Q} \left( g^{-+} - \frac{k^-n_s^+}{k^+n_s^- + k^-n_s^+} \right) 
\upperbub_\trans \pdfamp_\trans \right) \nonumber \\ \; \sim & \mathcal{O} \left( \frac{\Lambda}{Q} \upperbub_\trans \pdfamp_\trans \right) \label{eq:plusmin} \\
\upperbub^-\tilde{\pdfamp}^+ \sim & \mathcal{O} \left( \frac{Q}{\Lambda} \left( g^{+-} - \frac{k^+n_s^-}{k^+n_s^- + k^-n_s^+} \right) 
\upperbub_\trans \pdfamp_\trans \right) \nonumber \\ \; \sim & \mathcal{O} \left( \frac{\Lambda}{Q} e^{y_s} \upperbub_\trans \pdfamp_\trans \right). \label{eq:minplus}
\end{align}
Thus, after the replacement in Eq.~(\ref{eq:amp2}), the plus and minus 
components for a given graph are suppressed relative to the transverse components.  (Recall that $y_s \approx 0$.)
Note that it is important that the contribution from the $K$-term vanishes \emph{exactly} 
in the sum over all graphs in order to avoid 
uncontrolled errors, so we must use $k$ rather than $\tilde{k}$ in Eqs.~(\ref{eq:Gterm},~\ref{eq:Kterm}). 

As part of the wide-angle approximation,
we drop the power suppressed terms in Eqs.~(\ref{eq:plusmin},\ref{eq:minplus}) and 
make the replacement
\begin{multline}
\label{eq:amp3}
M(l_1,l_2,k) \to 
\sum_j \upperbub^{j}(l_1,l_2,k) 
\tilde{\pdfamp}_{j}(k,P) = \\ \sum_{i,j} \sum_s \upperbub^{j}(l_1,l_2,k)  \left( \epsilon_{t,j} \right)^s  \left( \epsilon_{t,i} \right)^s \tilde{\pdfamp}_{i}(k,P).
\end{multline}
The sums over $i$,$j$ are only over transverse momentum components.
In Eq.~(\ref{eq:amp3}) we have introduced the transverse polarization vectors
\begin{equation}
(\epsilon_{\trans , j})^1 = (0,1,0,0), \qquad (\epsilon_{\trans , j})^2 = (0,0,1,0).
\end{equation}
Using Eq.~(\ref{eq:amp3}) in the squared amplitude, summed over final states, we have
\begin{widetext}
\begin{equation} 
\label{eq:gluap1}
\left| M(l_1,l_2,k) \right|^2 \to \sum_{i,j,i^{\prime},j^{\prime}} \, \sum_{s,s^{\prime}} \; \left( \sum_{X_J} \upperbub^{j}(l_1,l_2,k) \, 
\upperbub^{j^\prime \, \dagger}(l_1,l_2,k) \right) \; 
 \left( \epsilon_{t,j} \right)^s  \left( \epsilon_{t,i} \right)^s \left( \epsilon_{t,j^\prime} \right)^{s^\prime}  \left( \epsilon_{t,i^\prime} \right)^{s^\prime} 
\left( \sum_{X_T} \tilde{\pdfamp}_{i}(k,P)  \tilde{\pdfamp}_{i^\prime}^{\dagger}(k,P) \right). 
\end{equation}
Here $\sum_{X_J}$ and $\sum_{X_T}$ are sums over the final states for the jets and the target respectively.
We can use the fact that, in an unpolarized cross section or structure function, $\upperbub^{j}(l_1,l_2,k) \, 
\upperbub^{j^\prime \, \dagger}(l_1,l_2,k)$ is 
diagonal in $j$ and $j^\prime$ to re-write Eq.~(\ref{eq:gluap1}) as
\begin{multline} 
\label{eq:glufactored}
\left( \left| M(l_1,l_2,k) \right|^2 \right)^{\mu \nu} \to  \left( \frac{1}{2} \sum_{X_J} \sum_{j} \, 
\upperbub^{j}(\approxa{l}_1,\approxa{l}_2,\approxa{k}) \, \upperbub^{j \, \dagger}(\approxa{l}_1,\approxa{l}_2,\approxa{k}) \right)^{\mu \nu} \; \
\left( \sum_{X_T} \sum_{j^\prime} G^{j^\prime \kappa} G^{j^\prime \kappa^\prime} \pdfamp_{\kappa}(k,P) \pdfamp_{\kappa^\prime}^{\dagger}(k,P) \right) \\
= \frac{1}{2} \sum_{s} \sum_{i,i^\prime} {\rm Tr} \left[\jetbub(l_1) \hardbub_L^{\mu, i}(l_1,l_2,k) \jetbub(l_2) \hardbub_R^{\nu, i^\prime}(l_1,l_2,k) \right] 
(\epsilon_{\trans, i})^{s} (\epsilon_{\trans, i^\prime})^{s}  \pdf_{g/p}^{(0)}(k,P).
\end{multline}
To symbolize the gluon projections described above, we use a projection matrix $\mathcal{P}^g_{j j^\prime}$,  
defined to implement the replacement in Eq.~(\ref{eq:amp3}).  That is, the replacement in Eq.~(\ref{eq:amp3}) is expressed as
\begin{equation}
M(l_1,l_2,k) \to \upperbub^j(l_1,l_2,k) \mathcal{P}^g_{j j^\prime} \pdfamp^{j^\prime}(k,P).
\end{equation}
\subsection{The fully unintegrated gluon PDF}

The square modulus of the amplitude in Fig.~\ref{fig:amplitude} is factored into two parts in Eq.~(\ref{eq:glufactored}).  The first factor is simply the 
on-shell squared matrix element for scattering off a transversely polarized gluon, averaged over transverse polarizations. 
The second factor
\begin{equation}
\label{eq:gluonpdf}
\pdf_{g/p}^{(0)}(k,P) = \sum_{X_T} \sum_{j^\prime} G^{j^\prime \kappa} G^{j^\prime \kappa^\prime} 
\pdfamp_{\kappa}(k,P) \pdfamp_{\kappa^\prime}^{\dagger}(k,P) = \\ \sum_{j^\prime} G^{j^\prime \kappa} G^{j^\prime \kappa^\prime} \Phi_{\kappa \kappa^\prime}(k,P) 
\end{equation}
is what we identify with the lowest order correction to the unpolarized fully unintegrated gluon PDF.
This suggests the following operator definition for the fully unintegrated gluon PDF:
\begin{equation}
\label{eq:gluedist}
\pdf_{g/p}(k,P) =
\sum_{j}^2 \int \frac{dw^+ dw^- d^{2}{\bf w}_\trans}{(2 \pi)^{4}}  
e^{-i k \cdot w} \langle P | \left[ n_{s , \mu} \fieldtensor^{\mu \, j}(w) \right]  \left[ n_{s , \nu} \fieldtensor^{\nu \, j}(0) \right] | P \rangle.
\end{equation} 
The sum over indices $j$ involves only the transverse components.
$\fieldtensor$ is the gauge field strength tensor.
The factorized structure of Eq.~(\ref{eq:glufactored})  
is shown 
graphically in Fig.~\ref{fig:ampfact} where the graph beneath the horizontal dotted line
now corresponds to Eq.~(\ref{eq:gluedist}), and the part above the 
dotted line corresponds to the first factor in parentheses in Eq.~(\ref{eq:glufactored}).
The cross drawn at the end of the target gluon is meant to symbolize that it is only the $G$-term that 
is retained in the target gluon propagator in Eq.~(\ref{eq:gluonpdf}).
In a non-Abelian gauge theory, Eq.~(\ref{eq:gluedist}) will need to be
modified by the insertion of a Wilson line operator, similar to the definition in Ref.~\cite{Collins:1981uw}, 
but these modifications won't effect the LO structure in Fig.~\ref{fig:ampfact}.
\begin{figure*}
\centering
    \includegraphics[scale=0.4]{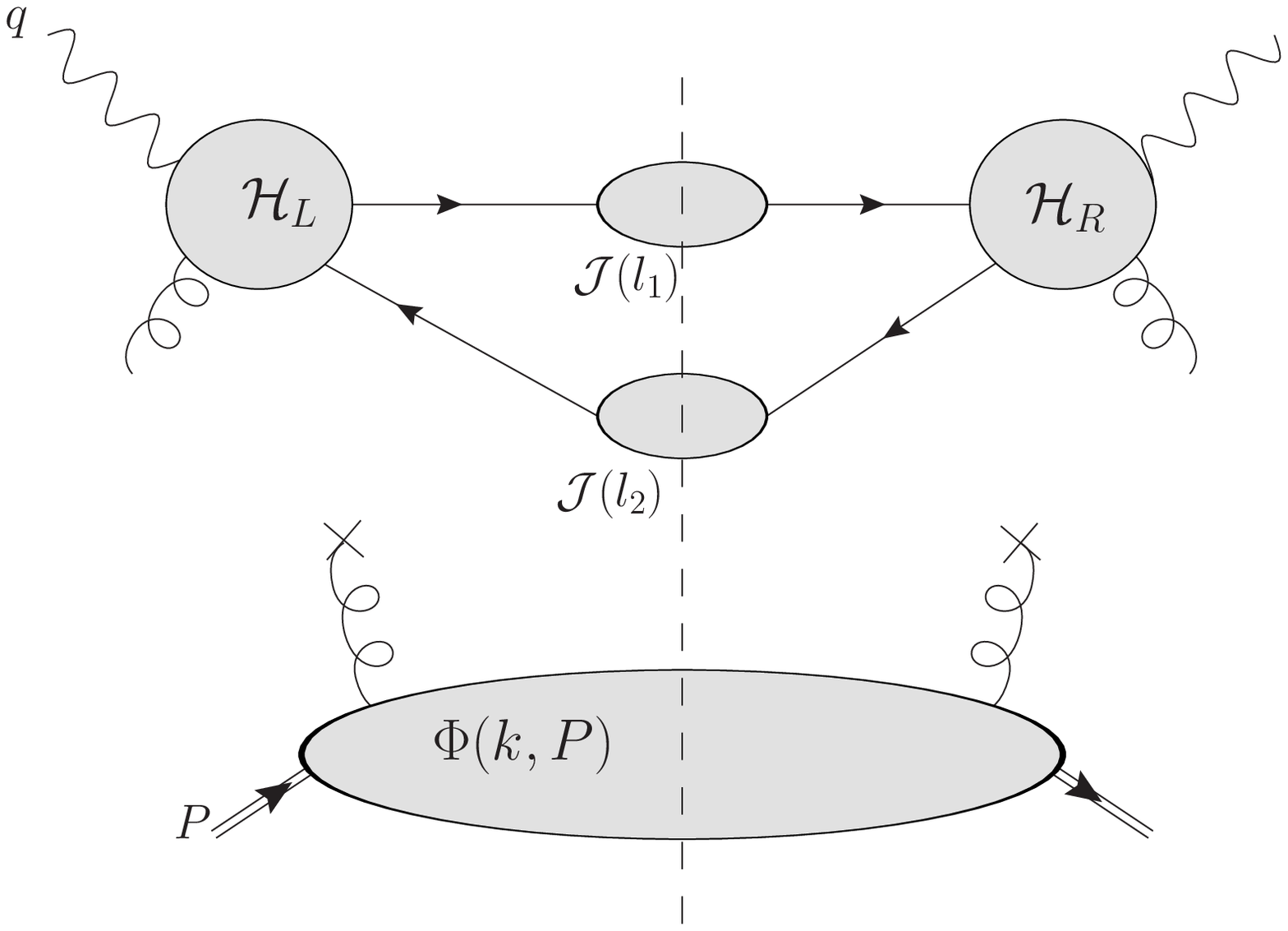}
\caption{The square modulus of the amplitude after keeping $G$-terms and projecting 
transverse components of the gluon propagator.  
The disconnected upper and lower parts represent the factors in Eq.~(\ref{eq:glufactored}).
The cross on the gluon in the lower bubble 
represents the contraction of the $G$-term with the lower bubble in Eq.~(\ref{eq:amp2}).}
\label{fig:ampfact}
\end{figure*}


\subsection{Final state jets}

Next we make projections onto relevant Dirac components for the 
outgoing jet lines, appropriate for region $R_2$. 
The final state jet bubbles can be expanded, as in Eq.~(\ref{eq:cliffordaa}), in a 
basis of the Dirac algebra.
The dominant components in these expansions are most easily found by working 
in coordinates where 
the $+z$ axis 
lies along the direction of $l_2$. 
Then $\approxa{l}_2$ has a large plus component and $\approxa{l}_1$ has a 
large minus component.  The dominant contributions for each jet bubble are then
\begin{align}
\label{eq:jetapproxs}
\gamma_\mu \jetbub^{\mu} (l_1) 
\approx
\frac{1}{(2 \approxa{s})^2} \slashed{\approxa{l}}_1 
\slashed{\approxa{l}}_2 \jetbub^{\mu}(l_1) \slashed{\approxa{l}}_2 \slashed{\approxa{l}}_1,  \\
\label{eq:jetapproxsb}
\gamma_\mu \jetbub^{\mu} (l_2) 
\approx
\frac{1}{(2 \approxa{s})^2} \slashed{\approxa{l}}_2 
\slashed{\approxa{l}}_1 \jetbub^{\mu}(l_2) \slashed{\approxa{l}}_1 \slashed{\approxa{l}}_2.
\end{align}
Since we restrict to unpolarized scattering, 
we have dropped polarization dependent terms.
In Eqs.~(\ref{eq:jetapproxs},\ref{eq:jetapproxsb}) we have used the 
approximate momenta to write the substitution in a frame independent way.
Now it is clear what the Dirac projection operators are for region $R_2$:
\begin{equation}
\label{eq:hardproj}
\mathcal{P}_{\approxa{l}_1} = \frac{1}{2 \approxa{s} } 
\slashed{\approxa{l}}_1 \slashed{\approxa{l}}_2, \qquad 
{\mathcal{P}}_{\approxa{l}_2} = \frac{1}{2 \approxa{s}} 
\slashed{\approxa{l}}_2 \slashed{\approxa{l}}_1. \\
\end{equation}
Part of the approximation for $R_2$ is then to make the substitutions
\begin{equation}
\jetbub(l_1) \to \mathcal{P}_{\approxa{l}_1} 
\jetbub(l_1) \mathcal{P}_{\approxa{l}_2}, \qquad \jetbub(l_2) 
\to \mathcal{P}_{\approxa{l}_2} \jetbub(l_2) \mathcal{P}_{\approxa{l}_1}.
\end{equation}
Equivalently, we may write
\begin{align}
\gamma_\mu \jetbub^{\mu} (l_1) 
&\approx  \frac{\slashed{\approxa{l}}_1}{2 \approxa{s}}  {\rm Tr} 
\left[ \slashed{\approxa{l}}_2 \jetbub(l_1) \right],\\
\gamma_\mu \jetbub^{\mu} (l_2) 
&\approx  \frac{\slashed{\approxa{l}}_2}{2 \approxa{s}}  {\rm Tr} 
\left[ \slashed{\approxa{l}}_1 \jetbub(l_2) \right].
\end{align} 
Making these substitutions in the squared amplitude, we have
\begin{multline}
\left( \left| M(l_1,l_2,k) \right|^2 \right)^{\mu \nu} 
\to {\rm Tr} \left[ \mathcal{P}_{\approxa{l}_1} 
\jetbub(l_1) \mathcal{P}_{\approxa{l}_2} \hardbub_L^{\mu, \kappa}(l_1,l_2,k) 
\mathcal{P}_{\approxa{l}_2} \jetbub(l_2) \mathcal{P}_{\approxa{l}_1} 
\hardbub_R^{\nu, \kappa^{\prime}}(l_1,l_2,k) \right] \pdfbub_{\kappa,\kappa^{\prime}}(k,P) \\
= 
{\rm Tr} \left[ \slashed{\approxa{l}}_1 \hardbub_L^{\mu, \kappa}(l_1,l_2,k) 
\slashed{\approxa{l}}_2 \hardbub_R^{\nu, \kappa^{\prime}}(l_1,l_2,k) 
\right]  
\left\{ \frac{1}{2 \approxa{s}} {\rm Tr} \left[ \slashed{\approxa{l}}_2 \jetbub(l_1) \right] \right\}
\left\{ \frac{1}{2 \approxa{s}} {\rm Tr} \left[ \slashed{\approxa{l}}_1 \jetbub(l_2) \right] \right\}
\pdfbub_{\kappa,\kappa^{\prime}}(k,P).
\label{eq:jet_replace}
\end{multline}
Finally, for convenience we define the following notation
\begin{eqnarray}
J_{\approxa{l}_2}(l_1) = \frac{1}{2 \approxa{s}} {\rm Tr} \left[ \slashed{\approxa{l}}_2 \jetbub(l_1) \right] = \picineq{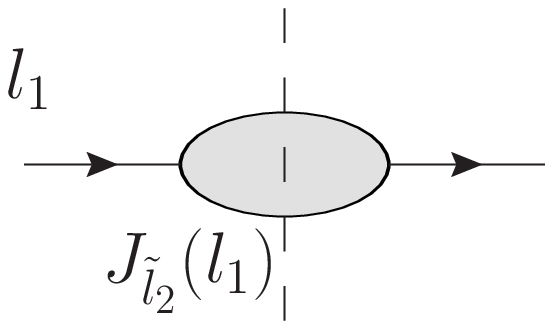} \\
J_{\approxa{l}_1}(l_2) = \frac{1}{2 \approxa{s}} {\rm Tr} \left[ \slashed{\approxa{l}}_1 \jetbub(l_2) \right] = \picineq{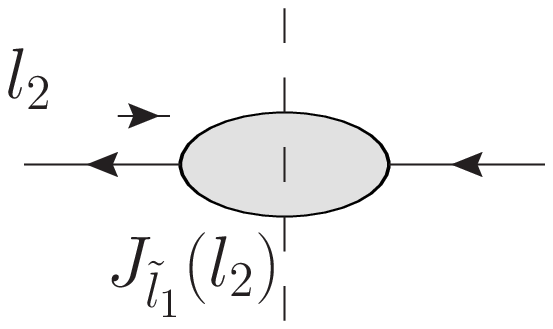}.
\end{eqnarray}
The subscript on $J$ indicates which vector is used for projecting Dirac indices.  Note the difference in normalization 
from Eqs.~(\ref{LOjet1},\ref{LOjet2}).
\subsection{Factorized expression for wide-angle $\gamma^\ast p \to 2 \, {\rm Jets} + X$ scattering}

We now have all the steps needed to factor the on-shell $\gamma^\ast g \to q \bar{q}$ matrix 
element from the rest of the graphs in Fig.~\ref{fig:boxdiagrams}.
Combining all of the approximations discussed in this section, appropriate for region $R_2$, amounts to 
replacing the hard scattering subgraphs by,
\begin{align} 
\hardbub_R^{\mu , \kappa}(l_1,l_2,k) \to 
\mathcal{P}_{\approxa{l}_2} \hardbub_R^{\mu, j}(\approxa{l}_1,\approxa{l}_2,\approxa{k})
\mathcal{P}_{\approxa{l}_2} \mathcal{P}^{g}_{j j^\prime}, \\ 
\hardbub_L^{\mu , \kappa}(l_1,l_2,k) \to 
\mathcal{P}^{g}_{i i^\prime}
\mathcal{P}_{\approxa{l}_1} \hardbub_L^{\mu , i}(\approxa{l}_1,\approxa{l}_2,\approxa{k})
\mathcal{P}_{\approxa{l}_1}. 
\end{align}
The approximator, $T_2$, appropriate for region $R_2$ acts on
the squared amplitude by making the complete set of replacements from this section in the graphs of Fig.~\ref{fig:boxdiagrams} and Eq.~(\ref{eq:decomp1}):
\begin{multline}
T_2 \left( \left| M(l_1,l_2,k) \right|^2 \right)^{\mu \nu} =
{\rm Tr} \left[ \mathcal{P}_{\approxa{l}_1} 
\jetbub(l_1) \mathcal{P}_{\approxa{l}_2} 
\hardbub_L^{\mu, j}(\approxa{l}_1,\approxa{l}_2,\approxa{k}) 
\mathcal{P}_{\approxa{l}_2} \jetbub(l_2) \mathcal{P}_{\approxa{l}_1} 
\hardbub_R^{\nu, i}(\approxa{l}_1,\approxa{l}_2,\approxa{k}) \right] 
 \mathcal{P}^{g}_{j j^\prime} \mathcal{P}^{g}_{i i^\prime}
\pdfbub^{j^{\prime} i^{\prime}}(k,P) \\ 
= \frac{1}{2} \sum_s \sum_{j j^\prime}
(\epsilon_{\trans, j})^{s} (\epsilon_{\trans, j^\prime})^{s} 
{\rm Tr} \left[ \slashed{\approxa{l}}_1 \hardbub_L^{\mu, j}(\approxa{l}_1,\approxa{l}_2,\approxa{k}) 
\slashed{\approxa{l}}_2 \hardbub_R^{\nu, j^\prime}(\approxa{l}_1,\approxa{l}_2,\approxa{k}) \right] 
\, J_{\approxa{l}_2}(l_1)
\, J_{\approxa{l}_1}(l_2)
\, \pdf_{g/p}(k,P) \\
= \left( \overline{\left| A^{\gamma^{\ast} g \rightarrow q \bar{q}}
(\approxa{l}_1,\approxa{l}_2,\approxa{k}) \right|}^2 \right)^{\mu \nu}
\, J_{\approxa{l}_2}(l_1)
\, J_{\approxa{l}_1}(l_2)
\, \pdf_{g/p}(k,P).
\label{eq:sqrdamp_ave3}
\end{multline}
In the second line we have repeated the steps that give Eqs.~(\ref{eq:glufactored},\ref{eq:jet_replace}), and we have dropped the 
small components in Eqs.~(\ref{eq:plusmin},\ref{eq:minplus}).  The ordinary unpolarized on-shell matrix element has been 
factored out of Eq.~(\ref{eq:sqrdamp_ave3}).

To summarize, Eq.~(\ref{eq:sqrdamp_ave3}) is the factorized expression for the graphs in Fig.~\ref{fig:boxdiagrams} that is 
a good approximation in the region of momentum space corresponding to two high transverse momentum jets.
The matrix element for massless on-shell $\gamma^{\ast} g \rightarrow q \bar{q}$ scattering with transversely polarized gluons 
has been factored out of the rest of the graph.  The last line in Eq.~(\ref{eq:sqrdamp_ave3}) 
can be interpreted diagrammatically as
\begin{multline}
\label{eq:regionh}
T_2 \left( \left| M(l_1,l_2,k) \right|^2 \right)^{\mu \nu} 
= \\
\left| \picineq{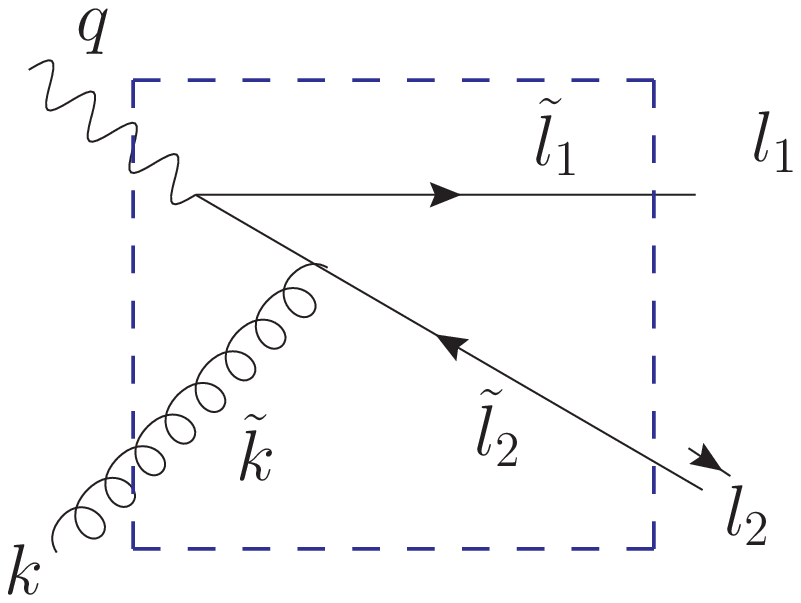} + \picineq{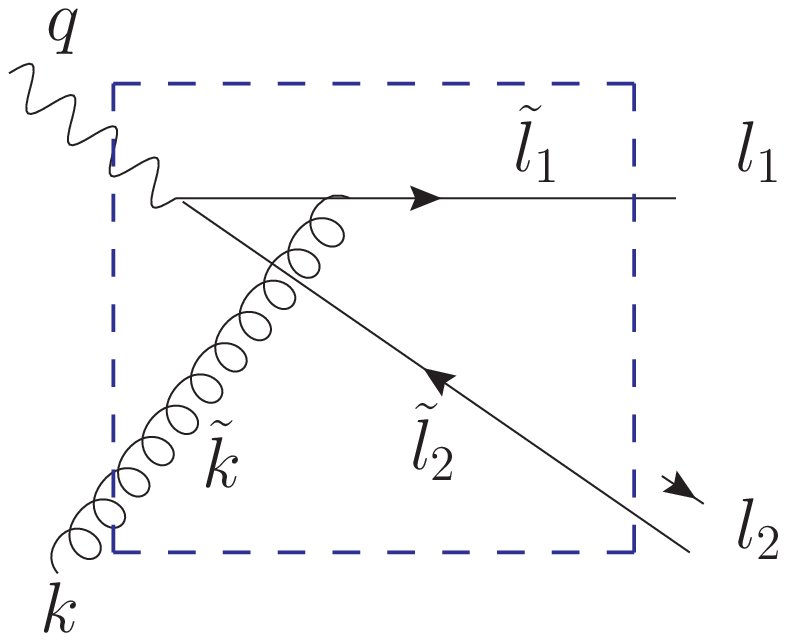} \right|^2 \times \left( \picineq{jetbub} \right) \times \left( \picineq{jetbub2} \right) \times \left( \picineq{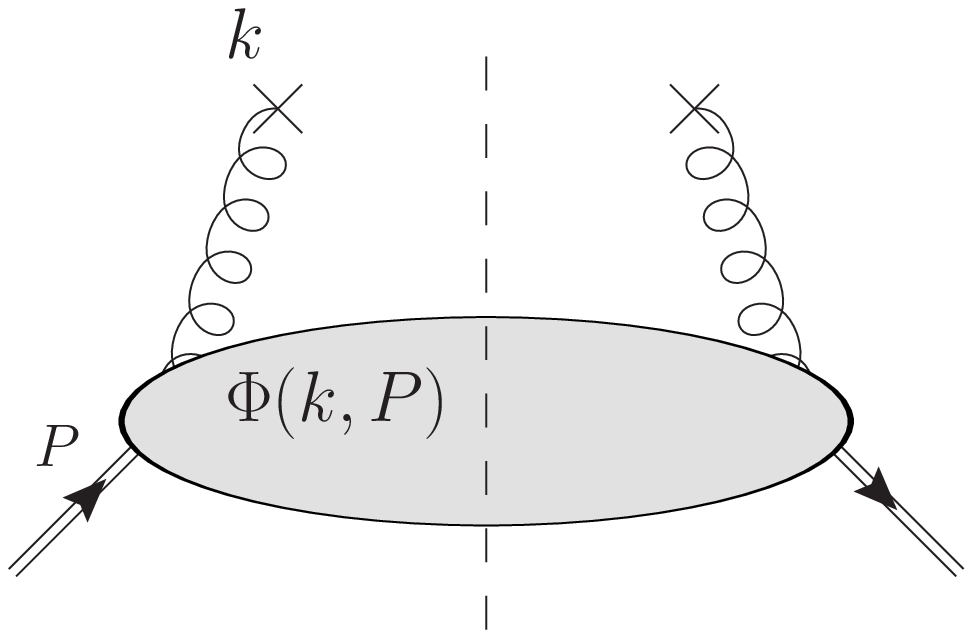} \right).
\end{multline}
To obtain the complete NLO result, we need to avoid 
double counting by subtracting from Eq.~(\ref{eq:sqrdamp_ave3}) the LO result of Sect.~\ref{sec:lowangle}.
This is the topic of the next section.

\section{Double Counting Subtractions}
\label{sec:double}

Recalling Eq.~(\ref{eq:wideangle}), the NLO contribution to the hadronic tensor is
\begin{equation}
\label{eq:fullsub}
\hadtensor^{\mu \nu}_{\gamma^\ast p \to q \bar{q}}(P,q) = T_2 \hadtensor^{\mu \nu}_g(P,q) - T_2 T_1 \hadtensor^{\mu \nu}_g(P,q) - T_2 \bar{T}_1 \hadtensor^{\mu \nu}_g(P,q).
\end{equation}
The first term in Eq.~(\ref{eq:fullsub}) is found by using the amplitude, Eq.~(\ref{eq:sqrdamp_ave3}), in Eq.~(\ref{eq:gammasummod}) 
for the structure tensor. 
The second term is the subtraction term corresponding to region $R_1$ and is found by directly evaluating
\begin{equation}
\label{eq:subterma}
T_2 T_1 \left( \left| M(l_1,l_2,k) \right|^2 \right)^{\mu \nu}.
\end{equation}
Similarly, the subtraction term for region $\bar{R}_1$ (target collinear anti-quark) is found by 
evaluating
\begin{equation}
\label{eq:subtermb}
T_2 \bar{T}_1 \left( \left| M(l_1,l_2,k) \right|^2 \right)^{\mu \nu}.
\end{equation}
With the careful formulation given so far of the approximations
for each region, it is now 
straightforward to obtain the correct subtraction term by 
directly applying the approximators, $T_1,\bar{T}_1,T_2$, in Eqs.~(\ref{eq:subterma},\ref{eq:subtermb}).
In this section we go through the steps of applying these approximations and we derive 
explicit expressions for the subtraction terms.
Returning again to Fig.~\ref{fig:boxdiagrams}, we consider each graph 
separately.

\subsection{Subtraction term for region $R_1$}
\subsubsection{Figure~\ref{fig:boxdiagrams}(a)}
\label{sec:subgrapha}

We start by calculating Eq.~(\ref{eq:subterma}),
beginning with the box diagram, Fig.~\ref{fig:boxdiagrams}a, and 
the corresponding amplitude Eq.~(\ref{eq:boxdigrama}).  
We first 
apply the approximator $T_1$ as it is defined in Sect.~\ref{sec:lowangle} and 
then the approximator $T_2$ as it is defined Sect~\ref{sec:hardregion}.
As in Sect.~\ref{sec:lowangle}, the application of the first approximator
is represented by solid red circles.  The application of the 
$T_2$ approximator is represented
by a dashed blue box enclosing the entire $\gamma^{\ast} g \rightarrow q \bar{q}$ 
block including the solid red circles (see Fig.~\ref{fig:bluered}).
Inside the dashed blue box, parton momenta are replaced by the 
approximate values in Eqs.~(\ref{eq:approxmoma},\ref{eq:approxmomab},\ref{eq:approxmomac}) (each denoted by a tilde).
Additionally, there is a projection at the intersection of each parton line 
with the dashed box.  Inside the solid red circle, parton momenta are
replaced by the approximate values in Eq.~(\ref{eq:core}) (each denoted by a hat).
Again, there are projections at each intersection of a parton line with the 
solid red circle.

The parton lines on either side of the electromagnetic 
vertices intersect the solid red circle so we make the replacements
\begin{equation}
\gamma^{\mu} \to \mathcal{P}_J \gamma^\mu \mathcal{P}_J, \qquad
\gamma^{\nu} \to \mathcal{P}_T \gamma^\nu \mathcal{P}_T,
\end{equation}
as prescribed by $T_1$.
Next, the external legs of the $\gamma^{\ast} g \rightarrow q \bar{q}$ subgraph
pass through the dashed blue lines, so we make the further replacements 
prescribed by $T_2$,
\begin{align}
\mathcal{P}_J \gamma^\mu \mathcal{P}_J \left( \frac{1}{\slashed{k}_1 - m} \right) \gamma^{\kappa} \to
\mathcal{P}^g_{j j^\prime} 
\mathcal{P}_{\approxa{l}_2} \mathcal{P}_J \gamma^{\mu} \mathcal{P}_J \left( \frac{1}{\slashed{\approxa{k}}_1} \right)
\gamma^{j} \mathcal{P}_{\approxa{l}_2} \\
\gamma^\rho \left( \frac{1}{\slashed{k}_1 - m} \right) \mathcal{P}_T \gamma^\nu \mathcal{P}_T  \to
\mathcal{P}_{\approxa{l}_1} \gamma^{i} \left( \frac{1}{\slashed{\approxa{k}}_1} \right)
\mathcal{P}_T \gamma^\nu \mathcal{P}_T \mathcal{P}_{\approxa{l}_1} 
\mathcal{P}^g_{i i^\prime}.
\end{align}
\begin{figure*}
\centering
  \begin{tabular}{c@{\hspace*{5mm}}c}
    \includegraphics[scale=0.4]{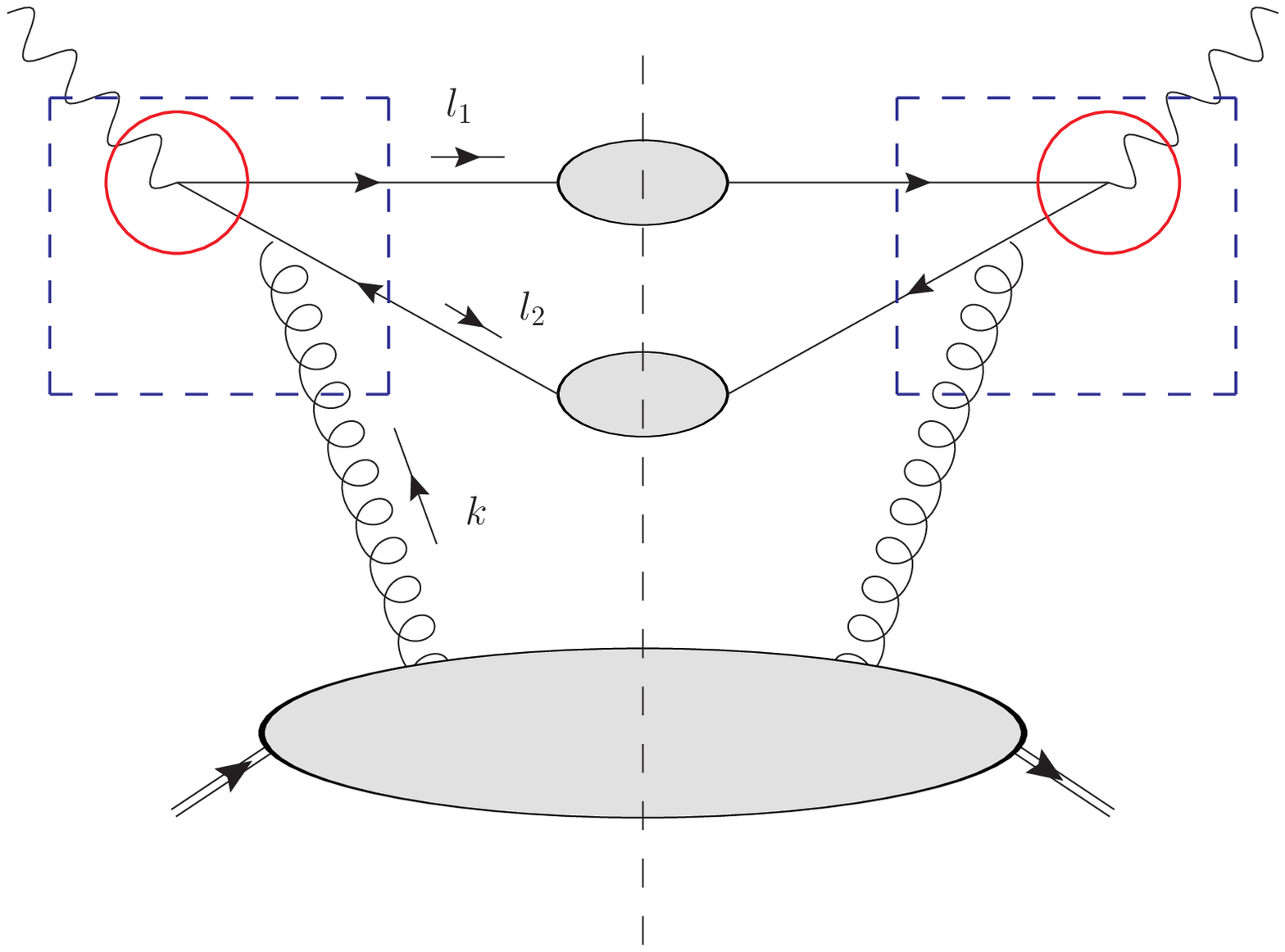}
    &
    \includegraphics[scale=0.4]{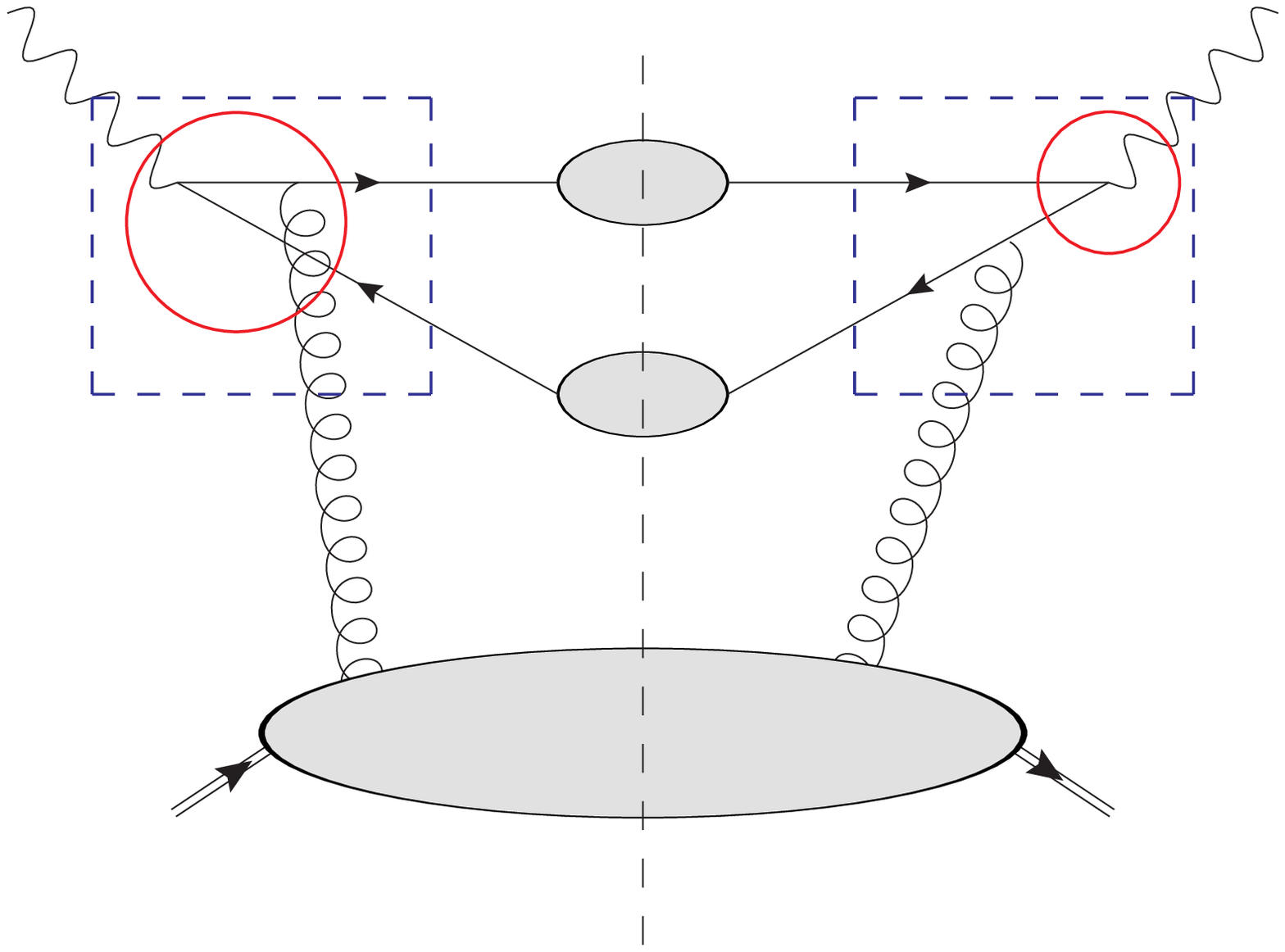}
  \\
    $(a)$ & $(b)$
  \end{tabular}
\caption{The sequence of approximations needed to obtain 
the subtraction terms corresponding to Figs.~\ref{fig:boxdiagrams}(a)-(b).  Analogous approximations are also need for Figs.~\ref{fig:boxdiagrams}(c)-(d) though 
we do not show them here.}
\label{fig:bluered}
\end{figure*}
There is a projection for each intersection of a 
parton line with the dashed blue boxes.  Additionally,
the momentum $k_1$ has been replaced by $\approxa{k}_1$ (Eq.~(\ref{eq:approxmoma})) 
inside the boxes and the masses of the internal partons are set to zero.  
Making these substitutions in the square modulus of the amplitude leads to
\begin{equation}
\label{eq:subandfactored}
T_2 T_1 \left( \left| M_a(l_1,l_2,k) \right|^2 \right)^{\mu \nu} 
= 
\strong T_R {\rm Tr} \left[ \gamma^{\nu} \mathcal{P}_T \mathcal{P}_{\approxa{l}_1}
\jetbub(l_1) \mathcal{P}_{\approxa{l}_2} \mathcal{P}_J \gamma^{\mu} \mathcal{P}_J \left( \frac{1}{\slashed{\approxa{k}}_1} \right)
\gamma^{j} \mathcal{P}_{\approxa{l}_2} \jetbub(l_2) 
 \mathcal{P}_{\approxa{l}_1} \gamma^{i} \left( \frac{1}{\slashed{\approxa{k}}_1} \right)
\mathcal{P}_T  \right] 
\mathcal{P}^g_{j j^\prime}
\mathcal{P}^g_{i i^\prime}
\pdfbub^{j^{\prime} i^{\prime}}(k,P).
\end{equation}
Repeating the steps of the last two sections allows Eq.~(\ref{eq:subandfactored}) to be written in a
more compact form with clearly recognizable factors:
\begin{multline}
\label{eq:subfactored}
T_2 T_1 \left( \left| M_a(l_1,l_2,k) \right|^2 \right)^{\mu \nu} 
=  \\
\frac{\strong T_R}{2 (2 \approxa{s})^2} \sum_s \sum_{i,j}
(\epsilon_{\trans, j})^{s} (\epsilon_{\trans, i})^{s} 
{\rm Tr} \left[ \gamma^{\nu} \mathcal{P}_T
\slashed{\approxa{l}}_1 \mathcal{P}_J \gamma^{\mu} \mathcal{P}_J \left( \frac{1}{\slashed{\approxa{k}}_1} \right)
\gamma^{j} \slashed{\approxa{l}}_2 \gamma^{i} \left( \frac{1}{\slashed{\approxa{k}}_1} \right)
\mathcal{P}_T \right] 
\left\{ 
{\rm Tr} \left[ \slashed{\approxa{l}}_2 \jetbub(l_1) \right] \right\}
\left\{ 
{\rm Tr} \left[ \slashed{\approxa{l}}_1 \jetbub(l_2) \right] \right\} 
\pdf_{g/p}(k,P) \\
=  
\frac{1}{2} {\rm Tr} \left[ \gamma^{\nu} 
\slashed{\approxb{l}}_1  \gamma^{\mu} \slashed{\approxb{k}}_1 \right] 
\fact_a(l_1,l_2,k) 
\left\{ \frac{1}{2 \approxa{s}} {\rm Tr} \left[ \slashed{\approxa{l}}_2 \jetbub(l_1) \right] \right\}
\left\{ \frac{1}{2 \approxa{s}} {\rm Tr} \left[ \slashed{\approxa{l}}_1 \jetbub(l_2) \right] \right\} 
\pdf_{g/p}(k,P),
\end{multline}
where in the second line we define the factor
\begin{equation}
\label{eq:facta}
\fact_a(l_1,l_2,k) = \frac{\strong T_R}{(2 Q^2)^2} \sum_s \sum_{i,j}
(\epsilon_{\trans, j})^{s} (\epsilon_{\trans, i})^{s}  {\rm Tr} 
\left[ \slashed{\approxb{k}}_1 \slashed{\approxa{l}}_1 \right] 
{\rm Tr} \left[ \slashed{\approxb{l}}_1 \left( \frac{1}{\slashed{\approxa{k}}_1} \right) \gamma^{j} 
\slashed{\approxa{l}}_2 \gamma^{i} \left( \frac{1}{\slashed{\approxa{k}}_1} \right) \right].
\end{equation}
Evaluating the contribution from Fig.~\ref{fig:blueboxes}(a) to Eq.~(\ref{eq:sqrdamp_ave3}) and then subtracting  Eq.~(\ref{eq:subfactored}) yields the contribution from 
Fig.~\ref{fig:boxdiagrams}(a) to the first two terms in Eq.~(\ref{eq:fullsub}):
\begin{multline}
\label{eq:suba}
T_{\rm 2} \left( \left| M_a(l_1,l_2,k) \right|^2 \right)^{\mu \nu}  - T_2 T_1 \left( \left| M_a(l_1,l_2,k) \right|^2 \right)^{\mu \nu} = \\
\left\{   \left( \overline{\left| A_a^{\gamma^{\ast} g \rightarrow q \bar{q}}
(\approxa{l}_1,\approxa{l}_2,\approxa{k}) \right|}^2 \right)^{\mu \nu}   - \fact_a(l_1,l_2,k) \left( \overline{\left| A^{\gamma^{\ast} q \rightarrow q}
(\approxb{l}_1,\approxb{k}_1) \right|}^2 \right)^{\mu \nu}  \right\} 
\, J_{\approxa{l}_2}(l_1)
\, J_{\approxa{l}_1}(l_2)
\, \pdf_{g/p}(k,P)
\end{multline}
One can verify, by repeating the steps above for $\bar{T}_1$, that Fig.~\ref{fig:boxdiagrams}(a) does not contribute to the third term in Eq.~(\ref{eq:fullsub}). 
Thus, the factor in braces in Eq.~(\ref{eq:suba}) is precisely contribution to the NLO fully unintegrated hard scattering coefficient 
arising from the application of the approximations and subtractions to Fig.~\ref{fig:boxdiagrams}(a).
Apart from $\fact_a(l_1,l_2,k)$ (defined in Eq.~(\ref{eq:facta})) Eq.~(\ref{eq:suba}) only involves ordinary on-shell 
squared amplitudes for parton scattering.
Graphically, the contribution to the hard scattering coefficient (in braces in Eq.~(\ref{eq:suba})) is
\begin{equation}
\label{eq:regionsuba}
\left| \picineq{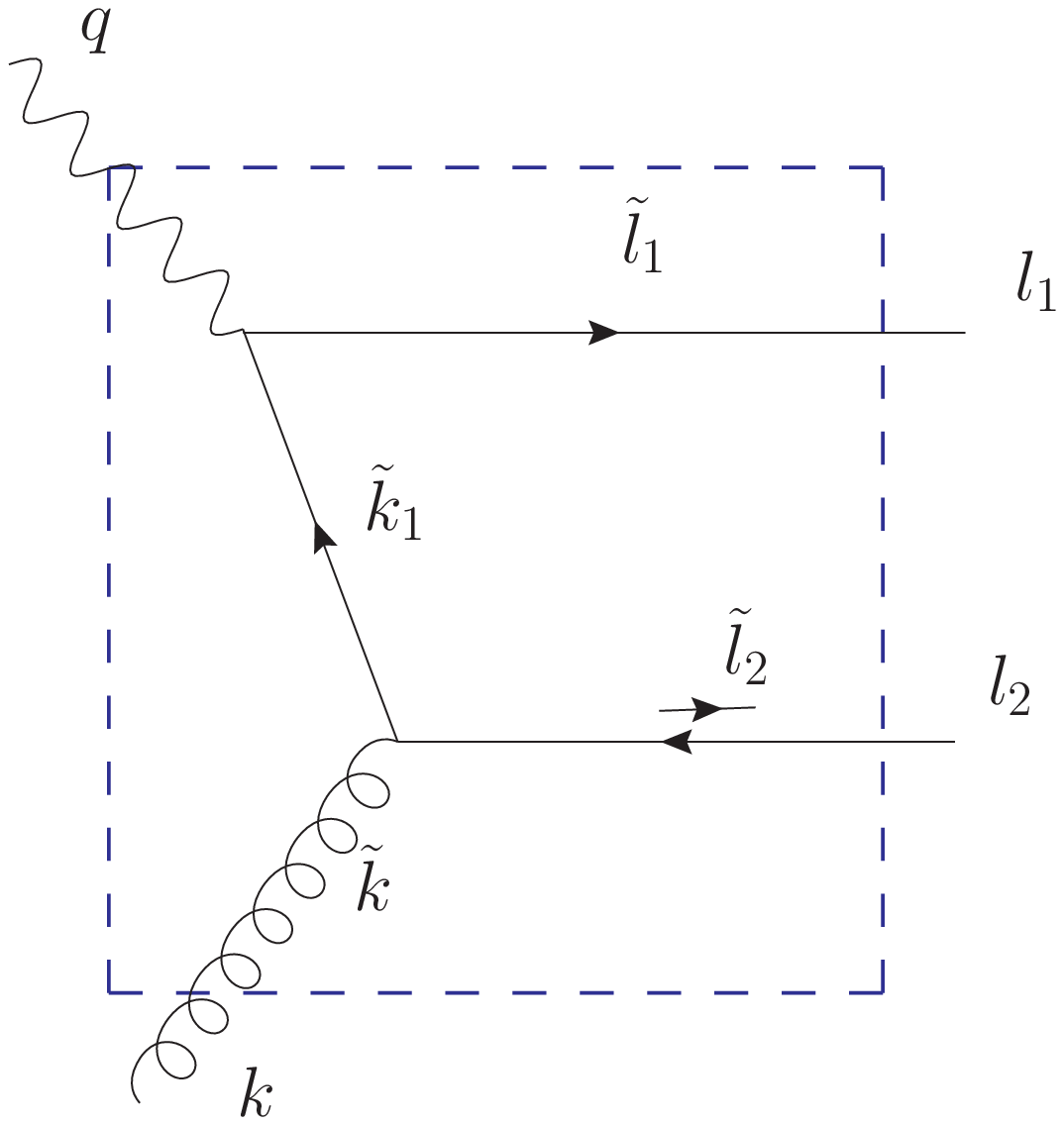}  \right|^2  \; \; \; - \; \; \; \left| \picineq{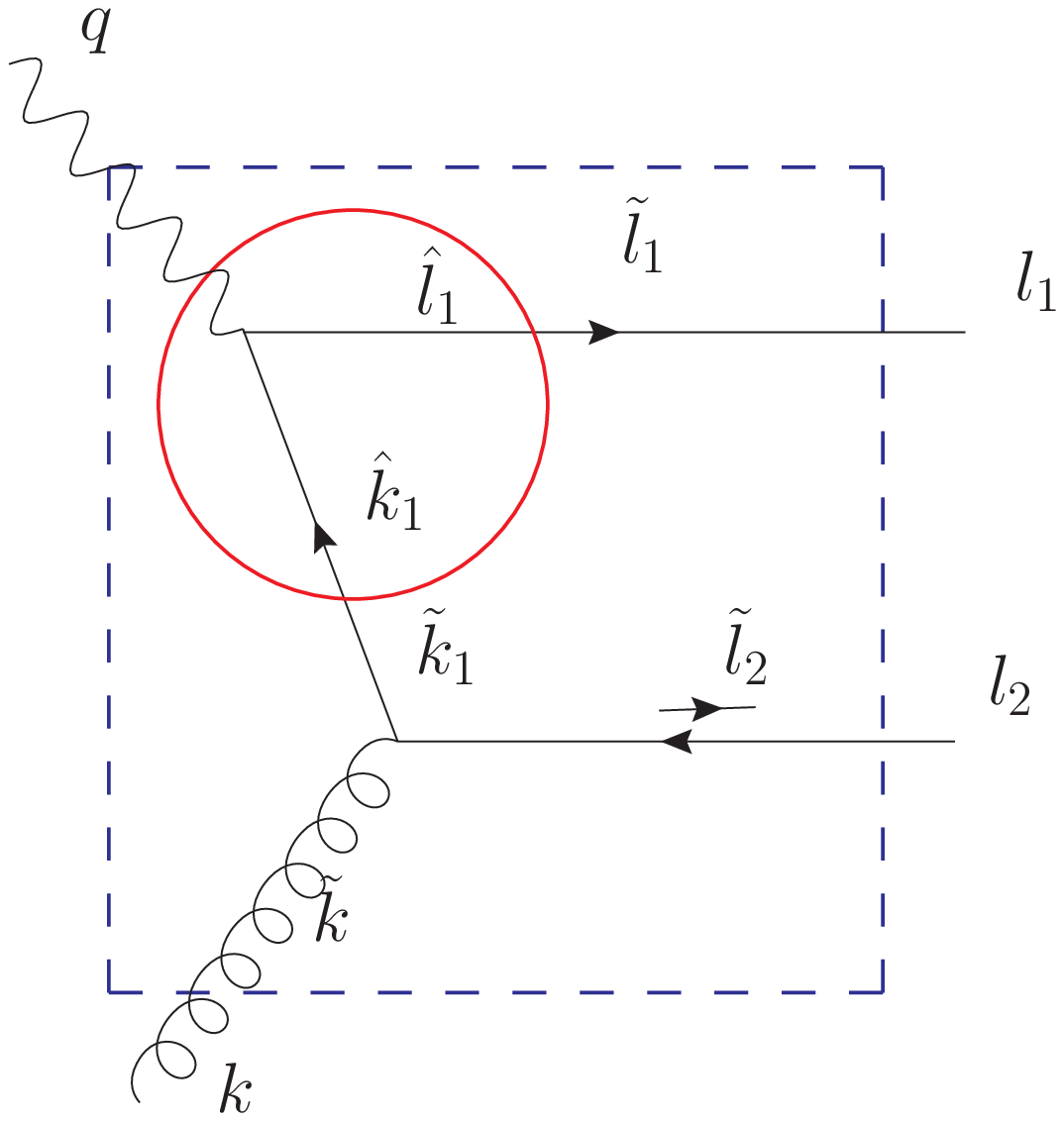} \right|^2.  
\end{equation}
The second factor here is the subtraction term.  The approximations that lead to Eq.~(\ref{eq:suba}) allow
the hard vertex to be factored out of the rest of the subtraction term.  Thus, we may write the graphical representation of the subtraction term in the simplified form:
\begin{equation}
\label{eq:vertfact}
\left| \picineq{subtractiona} \right|^2 \;\;\; = \;\;\; \fact_a(l_1,l_2,k) \left| \picineq{vertex} \right|^2.
\end{equation}
In the subtraction term, after the vertex factor is extracted as in Eq.~(\ref{eq:vertfact}), the singular behavior at $\approxa{k}_1 \to \approxb{k}_1$ 
is contained in the factor $\fact_a(l_1,l_2,k)$ which can be calculated directly by first writing it as
\begin{equation}
\label{eq:factmod}
\fact_a(l_1,l_2,k) = \frac{\strong T_R}{(2 Q^2)^2} \; g^t_{\kappa \rho}
{\rm Tr} 
\left[ \slashed{\approxb{k}}_1 \slashed{\approxa{l}}_1 \right] 
{\rm Tr} \left[ \slashed{\approxb{l}}_1 \left( \frac{1}{\slashed{\approxa{k}}_1} \right) \gamma^{\kappa} 
\slashed{\approxa{l}}_2 \gamma^{\rho} \left( \frac{1}{\slashed{\approxa{k}}_1} \right) \right],
\end{equation}
where we define the transverse tensor
\begin{equation}
\label{eq:gtrans}
g^t_{\kappa \rho} = - g_{\kappa \rho} + \frac{\approxa{k}_\kappa n_{J,\rho} + n_{J,\kappa} \approxa{k}_\rho}{\approxa{k} \cdot n_J}, \qquad n_J = (0,1,{\bf 0}_t).
\end{equation}
Evaluating Eq.~(\ref{eq:factmod}) yields
\begin{equation}
\label{eq:subfact}
\fact_a(l_1,l_2,k) = -\frac{\strong T_R}{Q^4 \approxa{t}} \; (2 \approxb{k}_1 \cdot \approxa{l}_1 ) \left( 4 \approxb{l}_1 \cdot \approxa{k} 
- \frac{(2 \approxb{l}_1 \cdot \approxa{k}_1) (2 \approxa{l}_2 \cdot n_J)}{\approxa{k} \cdot n_J} - \frac{(2 \approxb{l}_1 \cdot \approxa{l}_2) (2 \approxa{k}_1 \cdot n_J)}{\approxa{k} \cdot n_J} \right). 
\end{equation}
Note that $\fact_a$ depends on the exact momenta, $l_1$,$l_2$, and $k$, via 
the dependence of the approximate momentum variables on exact momenta as defined in Eq.~(\ref{eq:core},\ref{eq:approxmoma}-\ref{eq:approxmomac}).

\subsubsection{Figures~\ref{fig:boxdiagrams}(b)-(d)}
\label{sec:subgraphb}
We may also go through the steps for Fig.~\ref{fig:boxdiagrams}(b).
The solid red circles and 
blue dashed boxes denoting the sequence of approximations are shown in Fig.~\ref{fig:bluered}(b).
Starting again with Eq.~(\ref{eq:ampunapproxb}), 
we first apply the approximator for region $R_1$.  The red circles tell us to make the
replacements
\begin{equation}
\label{eq:pmapproxb}
\gamma^{\kappa} \left( \frac{1}{\slashed{l}_1 - \slashed{k} - m} \right) \gamma^{\mu} \to \mathcal{P}_J 
\gamma^{\kappa} \left( \frac{1}{\slashed{\approxb{l}}_1 - \slashed{\approxb{k}}} \right) \gamma^{\mu} \mathcal{P}_J, \qquad
\gamma^{\nu} \to \mathcal{P}_T \gamma^\nu \mathcal{P}_T.
\end{equation}
Additionally, in accordance with our prescription for region $R_1$, we dot $\slashed{\approxb{k}}$ into 
the left-hand circle and multiply by $n_s^{\kappa}/(n_s \cdot k)$.  
So we make the further substitution
\begin{equation}
\label{eq:pmapproxb2}
\mathcal{P}_J 
\gamma^{\kappa} \left( \frac{1}{\slashed{\approxb{l}}_1 - \slashed{\approxb{k}}} \right) \gamma^{\mu} \mathcal{P}_J \to \frac{n_s^{\kappa}}{n_s \cdot k} \mathcal{P}_J 
\slashed{\approxb{k}} \left( \frac{1}{\slashed{\approxb{l}}_1 - \slashed{\approxb{k}}} \right) \gamma^{\mu} \mathcal{P}_J.
\end{equation}
The outer rectangle tells us to apply the approximations for region $R_2$ to everything inside it.
Therefore, on the left-hand side we make the substitutions
\begin{equation}
\label{eq:subgraphb}
\frac{n_s^{\kappa}}{n_s \cdot k} \mathcal{P}_J 
\slashed{\approxb{k}} \left( \frac{1}{\slashed{\approxb{l}}_1 - \slashed{\approxb{k}}} \right) \gamma^{\mu} \mathcal{P}_J \to
\mathcal{P}^g_{j j^\prime} \mathcal{P}_{l_2} \frac{n_s^j}{n_s \cdot \approxa{k}} \mathcal{P}_J 
\slashed{\approxb{k}} \left( \frac{1}{\slashed{\approxb{l}}_1 - \slashed{\approxb{k}}} \right) \gamma^{\mu} \mathcal{P}_J \mathcal{P}_{l_2} = 0.
\end{equation}
The left-hand hard subgraph in graph (b) exactly vanishes because $\mathcal{P}^g_{j j^\prime}$ involves 
only transverse components (see Eq.~\ref{eq:amp3}) whereas $n_s$ has no transverse components.
A similar result holds for graphs (c) and (d).  Thus, only Eq.~(\ref{eq:suba}) 
contributes to the region $R_1$ subtraction term.

\subsection{Subtraction term for region $\bar{R}_1$}
The steps for calculating the anti-quark subtraction term, Eq.~(\ref{eq:subtermb}), 
follow exactly analogous steps to what was already done in the previous subsection for region $R_1$, so we 
do not repeat the details.  Instead of Fig.~\ref{fig:boxdiagrams}(a), it 
is Fig.~\ref{fig:boxdiagrams}(d) that gives the only non-vanishing subtraction term 
in region $\bar{R}_1$.  The analogue of Eq.~(\ref{eq:suba}) is
 \begin{multline}
\label{eq:subd}
T_{\rm 2} \left( \left| M_d(l_1,l_2,k) \right|^2 \right)^{\mu \nu}  - T_2 \bar{T}_1 \left( \left| M_d(l_1,l_2,k) \right|^2 \right)^{\mu \nu} = \\
\left\{   \left( \left| A_d^{\gamma^{\ast} g \rightarrow q \bar{q}}
(\approxa{l}_1,\approxa{l}_2,\approxa{k}) \right|^2 \right)^{\mu \nu}   - \fact_d(l_1,l_2,k) \left( \overline{\left| A^{\gamma^{\ast} \bar{q} \rightarrow \bar{q}}
(\check{l}_2,\check{k}_1^\prime) \right|}^2 \right)^{\mu \nu}  \right\} 
\, J_{\approxa{l}_2}(l_1)
\, J_{\approxa{l}_1}(l_2)
\, \pdf_{g/p}(k,P).
\end{multline}
The factor in the subtraction term analogous to Eq.~(\ref{eq:subfact}) is
\begin{equation}
\label{eq:subfactbar}
\bar{\fact}_d(l_1,l_2,k) = -\frac{\strong T_R}{Q^4 \approxa{u}} \; (2 \check{k}_1^\prime \cdot \approxa{l}_2 ) \left( 4 \check{l}_2 \cdot \approxa{k} 
- \frac{(2 \check{l}_2 \cdot \approxa{k}_1^\prime) (2 \approxa{l}_1 \cdot n_J)}{\approxa{k} \cdot n_J} - \frac{(2 \check{l}_2 \cdot \approxa{l}_1)(2 \approxa{k}_1^\prime \cdot n_J)}{\approxa{k} \cdot n_J} \right). 
\end{equation}
\section{Complete NLO Contribution}
\label{sec:NLO}
Gathering the results of the previous section together, we may write 
down the order-$\strong$ contribution to Eq.~(\ref{eq:gammasummod}):
\begin{multline}
\label{eq:gammasummodb}
\hadtensor_{\gamma^\ast g \to q \bar{q}}^{\mu \nu}(P,q) = \\ = \frac{e_j^2}{4 \pi} \int \frac{d^4 l_{2}}{(2 \pi)^4 } 
\int \frac{d^4 l_{1}}{(2 \pi)^4 } \int \frac{d^4 k}{(2 \pi)^4 } 
\tilde{W}_{\gamma^{\ast} g \to q \bar{q}}^{\mu \nu}(l_1,l_2,k)
\, J_{\approxa{l}_2}(l_1)
\, J_{\approxa{l}_1}(l_2)
\, \pdf_{g/p}(k,P) \,
(2 \pi)^4 \delta^{(4)}(k + q - l_1 - l_2).
\end{multline}
This is the desired factorized form for the third term in the last line of Eq.~(\ref{eq:wideangle}).
The fully unintegrated hard scattering coefficient is, from Eqs.~(\ref{eq:suba},\ref{eq:subd}),
\begin{multline}
\label{eq:fullysubtracted}
\tilde{W}_{\gamma^{\ast} g \to q \bar{q}}^{\mu \nu}(l_1,l_2,k) = \\
\left( \overline{\left| A^{\gamma^{\ast} g \rightarrow q \bar{q}}(\approxa{l}_1,\approxa{l}_2,\approxa{k}) \right|}^2 \right)^{\mu \nu}   
- \fact_a(l_1,l_2,k)  \left( \overline{\left| A^{\gamma^{\ast} q \rightarrow q}(\approxb{l}_1,\approxb{k}_1) \right|}^2 \right)^{\mu \nu}
- \bar{\fact}_d(l_1,l_2,k)  \left( \overline{\left| A^{\gamma^{\ast} \bar{q} \rightarrow \bar{q}}(\check{l}_2,\check{k}_1^\prime) \right|}^2 \right)^{\mu \nu}.  
\end{multline}
In terms of diagrams, the hard scattering coefficient is
\begin{multline}
\label{eq:Wgraph}
\tilde{W}_{\gamma^{\ast} g \to q \bar{q}}^{\mu \nu}(l_1,l_2,k)
= 
\left| \picineq{gpqq_amp_a} + \picineq{gpqq_amp_b} \right|^2  -  
\\ 
- \fact_a(l_1,l_2,k) \left| \picineq{vertex} \right|^2 
- \bar{\fact}_d(l_1,l_2,k) \left| \picineq{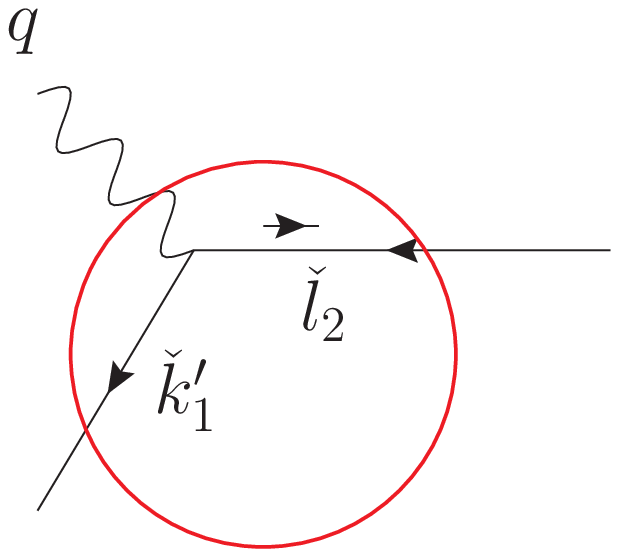} \right|^2. 
\end{multline}
The steps for evaluating the integral in Eq.~(\ref{eq:gammasummodb}) are 
as follows:
\begin{itemize}
\item Obtain values of $l_1$, $l_2$ and $k$ directly from parameterizations or models of $J_{\approxa{l}_2}(l_1)$, $J_{\approxa{l}_1}(l_2)$, and $\pdf_{g/p}(k,P)$.
\item Determine the correct approximate values of parton momentum ($\approxb{l}_1$,$\approxb{k}_1$,$\check{l}_2$,$\check{k}_1^\prime$,$\approxa{l}_1$,$\approxa{l}_2$,$\approxa{k}$) 
using Eqs.~(\ref{eq:core},\ref{eq:corebar},\ref{eq:approxmoma}-\ref{eq:approxmomac}).
\item Evaluate Eq.~(\ref{eq:fullysubtracted}) using these approximate momenta.  The factors, $\fact_a(l_1,l_2,k)$ and $\bar{\fact}_d(l_1,l_2,k)$
are to be directly evaluated using Eqs.~(\ref{eq:subfact},\ref{eq:subfactbar}).  The squared amplitudes 
in Eq.~(\ref{eq:Wgraph})
are just the usual ones obtained using Feynman graphs, but evaluated with 
the appropriate approximate parton momentum.  Calculating these amplitudes requires that we make a projection on electromagnetic indices 
to obtain a particular structure function.  For completeness, we give explicit expressions for the squared partonic amplitudes, with projections on electromagnetic indices, in App.~\ref{partamp}.
\end{itemize}

The first term in Eq.~(\ref{eq:fullysubtracted}) contains singularities at $\approxa{k}_1 \to \approxb{k}_1$ and $\approxa{k}_1^\prime \to \check{k}_1^\prime$
which are exactly canceled by the subtraction terms where the singularities are contained in the factors, $\fact_a(l_1,l_2,k)$ and $\bar{\fact}_d(l_1,l_2,k)$. 
This is similar to what occurs in the 
usual subtraction approach to factorization.  
However, in the standard integrated approach to factorization, the subtractions 
involve generalized functions (e.g., $\delta$-functions and ``$+$''-distributions), so that 
the cancellation only makes sense \emph{after} an integration over final states.
By contrast, the fully unintegrated hard scattering coefficient Eq.~(\ref{eq:fullysubtracted}) is just an ordinary function that 
can be evaluated for any final state momentum.  
The subtractions of Ref.~\cite{CZ,CRS} that ensure consistent factorization occur point-by-point in momentum space. 
Using Eq.~(\ref{eq:gmunuLO}-\ref{eq:pmunuNLO}) in App.~\ref{partamp}, we have verified numerically that the cancellation of singularities takes place.
\end{widetext}

\section{Summary and Conclusion}
\label{eq:complete}

With the results of this paper we are able to directly calculate   
the gluon-induced contribution to the unpolarized structure functions to order $\strong$ in 
a fully unintegrated approach.
The key results are the LO contributions, Eqs.~(\ref{eq:T1},\ref{eq:T1bar}), and 
the NLO contribution Eq.~(\ref{eq:gammasummodb}) to the hadronic structure tensor, Eq.~(\ref{eq:gammasummod}).
The LO and NLO results should be summed to accurately describe the overlap of 
the different regions of momentum space.
Since we have used the subtractive method of Refs.~\cite{CZ,JCC}, then we can be confident that the 
errors are $\Lambda/Q$ suppressed point-by-point in momentum space.

Evaluating Eqs.~(\ref{eq:T1},\ref{eq:T1bar},\ref{eq:gammasummod}) only requires the expressions for the 
matrix elements obtained from ordinary Feynman graph methods (see Eqs.~(\ref{eq:gmunuLO})-(\ref{eq:pmunuNLO})), and the explicit expressions given for the factors 
$\fact_a(l_1,l_2,k)$ and $\bar{\fact}_d(l_1,l_2,k)$ in Eqs.~(\ref{eq:subfact},\ref{eq:subfactbar}).
Finally, the approximate variables used to evaluate the different factors (\ref{eq:gmunuLO}-\ref{eq:pmunuNLO},\ref{eq:subfact},\ref{eq:subfactbar}) are given by the mappings in
Eqs.~(\ref{eq:core},\ref{eq:approxb2},\ref{eq:approxmoma}-\ref{eq:approxman}) which relate exact to approximate momentum variables. 

Since there is a large overlap of regions $R_1$, $\bar{R}_1$ with $R_2$, the subtraction terms need to be dealt with very carefully.
As an example of how things could go wrong,   
we could imagine that we approximate $\approxa{k}$ in the numerator of the subtraction factors, Eqs.~(\ref{eq:subfact},\ref{eq:subfactbar}), by
$\approxb{k}$ from Eq.~(\ref{eq:approxb2}).  
This could simplify the expressions and, naively, it might seem to make sense given that the purpose of the 
subtraction terms is to remove double counting from $R_1$ and $\bar{R}_1$ where $\approxb{k}$ is a good approximation to $k$.
However, the subtraction terms are \emph{used in regions of momentum space that are far from the core of the $R_1$ and $\bar{R}_1$ regions}.
Thus, replacing $\approxa{k}$ by $\approxb{k}$ in Eqs.~(\ref{eq:subfact},\ref{eq:subfactbar})
yields large modifications to 
Eq.~(\ref{eq:Wgraph})
in certain 
regions of phase space.
Alternatively, one could imagine trying to change the definition of $\approxb{k}$ in Eq.~(\ref{eq:approxb2}).  However, 
the definition of $\approxb{k}$ that we have used is rigidly fixed by the requirements of the factorization formula in Ref.~\cite{CRS}, with a 
well-defined fully unintegrated quark PDF.  We can thus see how the lowest order factorization formula 
and the corresponding 
definitions for the fully unintegrated PDFs places important restrictions on the higher order calculations. 

We have so far considered only the $\gamma^\ast g \to q \bar{q}$ partonic subprocess. 
To complete the NLO treatment requires that we include the somewhat more complicated
$\gamma^\ast q \to g q$ partonic subprocess, so the result of this paper is not quite a complete 
NLO treatment.
However, this paper provides a concrete illustration of the techniques developed in Ref.~\cite{CZ,CRS} applied to order-$\strong$ scattering, and 
provides a starting point for considering more complicated diagrams.
Furthermore, the graphs in Fig.~\ref{fig:boxdiagrams} dominate processes at low-$x$ and are the only graphs included
in some MCEGs.  Thus, the results presented here are already of phenomenological interest.
The treatment in this paper can be implemented directly in Monte Carlo calculations 
similar to what was done in Ref.~\cite{Collins:2005uv,CASCADE} where it was shown that fully unintegrated PDFs are needed 
to obtain an accurate description of the distribution in final states.
Using the results of this paper, one has direct access to the fully unintegrated gluon PDF.
Furthermore, the techniques described here are likely to be 
compatible with existing methods (\cite{MRW1,MRW2,Hoche:2007hg,Ciafaloni:1987ur,Catani:1989sg,GolecBiernat:2007pu,Gieseke:2007ad}) in the description of initial and final states;
a description of the non-perturbative bubbles, $\Phi$ and $\jetbub$, can likely be obtained by utilizing these other approaches.

In the large-$x$ limit, the standard kinematic approximations of collinear factorization can affect even totally inclusive quantities, and 
a fully unintegrated formalism may be needed.
The set-up in this paper naturally takes large-$x$ kinematic effects into account.
An alternative, simpler approach was developed in Ref.~\cite{Accardi:2008ne} where 
cross sections where written in terms of the usual integrated PDFs of the standard 
collinear factorization theorems, but such that large-$x$ effects from initial and final state masses were taken into account. 
It is possible that this approach can be extended to the order-$\strong$ jet production calculation presented in this paper in the large-$x$ limit, 
thus simplifying the evaluation of Eq.~(\ref{eq:gammasummodb}).
(In the large-$x$ limit, however, it is not sufficient to keep only the diagrams 
of Fig.~\ref{fig:boxdiagrams}, and quark-induced processes also need to be included.)

The natural next step is to extend the calculation illustrated in this paper to the case of 
single gluon emission.  That is, for the partonic subprocess, we also need to consider diagrams 
for $\gamma^\ast q \to g q$ scattering.  The calculation will follow steps very similar to those presented here.
However, it will be somewhat complicated by the need to take into account subtractions that correspond
to contributions to the soft factor and the jet factor.  Therefore, the soft factor will need 
to be explicitly included everywhere.

\section*{Acknowledgments} 
I would like to thank my colleagues, John Collins, Hannes Jung, Anna Sta\'sto, and Mark Strikman for 
many very useful discussions.

Feynman diagrams 
were made using JaxoDraw~\cite{Binosi:2003yf}.
This work was supported by the
U.S. D.O.E. under grant number DE-FG02-90ER-40577.

\begin{widetext}
\appendix
\section{Partonic Amplitudes}
\label{partamp}
Structure functions like Eq.~(\ref{eq:F1}) are calculated by making projections on electromagnetic indices using suitable combinations of $g^{\mu \nu}$ and $P^\mu P^\nu$.
Here we write down the explicit expressions obtained by projecting  with $g^{\mu \nu}$ and $P^\mu P^\nu$ on the partonic squared amplitudes in 
Eq.~(\ref{eq:fullysubtracted}).
For the LO partonic amplitude the result is
\begin{eqnarray}
g_{\mu \nu} \left( \overline{\left| A^{\gamma^{\ast} q \rightarrow q} (\approxb{l}_1,\approxb{k}_1) \right|}^2 \right)^{\mu \nu}
& = &  -2 Q^2, \label{eq:gmunuLO} \\
\, \nonumber \\
P_{\mu} P_{\nu} \left( \overline{\left| A^{\gamma^{\ast} q \rightarrow q} (\approxb{l}_1,\approxb{k}_1) \right|}^2 \right)^{\mu \nu} 
& = & 0. \label{eq:pmunuLO} \\
\, \nonumber 
\end{eqnarray}
\begin{eqnarray}
g_{\mu \nu} \left( \overline{\left| A^{\gamma^{\ast} q \rightarrow q} (\check{l}_2,\check{k}_1^\prime) \right|}^2 \right)^{\mu \nu}
& = &  -2 Q^2, \label{eq:gmunuLObar} \\
\, \nonumber \\
P_{\mu} P_{\nu} \left( \overline{\left| A^{\gamma^{\ast} q \rightarrow q} (\check{l}_2,\check{k}_1^\prime) \right|}^2 \right)^{\mu \nu} 
& = & 0. \label{eq:pmunuLObar} \\
\, \nonumber 
\end{eqnarray}
For the NLO partonic amplitude the result is
\begin{equation}
g_{\mu \nu} \left( \overline{\left| A^{\gamma^{\ast} g \rightarrow q \bar{q}}(\approxa{l}_1,\approxa{l}_2,\approxa{k}) \right|}^2 \right)^{\mu \nu} 
 =  - 4 \strong T_R \left( \frac{\approxa{t}}{\approxa{u}} + \frac{\approxa{u}}{\approxa{t}} - \frac{2 \approxa{s} Q^2}{\approxa{t} \approxa{u}} \right), \label{eq:gmunuNLO} 
\end{equation}
and,
\begin{multline}
P_{\mu} P_{\nu} \left( \overline{\left| A^{\gamma^{\ast} g \rightarrow q \bar{q}}(\approxa{l}_1,\approxa{l}_2,\approxa{k}) \right|}^2 \right)^{\mu \nu} 
 = 
- 2 \strong T_R  
\left( (2 \approxa{k} \cdot P) \left( \frac{2 \approxa{l}_1 \cdot P}{\approxa{t}} + \frac{2 \approxa{l}_2 \cdot P}{\approxa{u}}  \right) 
+ M_p^2 \left( \frac{\approxa{u}}{\approxa{t}} + \frac{\approxa{t}}{\approxa{u}}  \right) \right) + \\ 
 +  \frac{2\strong T_R}{\approxa{t} \approxa{u}} \left( 2 \approxa{s} Q^2 M_p^2 - (2 P \cdot \approxa{l}_1) ( \, (2 P \cdot \approxa{k}_1)(2 \approxa{k}_1^\prime \cdot \approxa{l}_2) + \approxa{t} \, (2 P \cdot \approxa{k}_1^\prime) \, ) 
- 
(2 P \cdot \approxa{l}_2) ( \, (2 P \cdot \approxa{k}_1^\prime)(2 \approxa{k}_1 \cdot \approxa{l}_1) - \approxa{u} \, (2 P \cdot \approxa{k}_1) \, ) \right).
\label{eq:pmunuNLO}
\end{multline}
We have verified numerically that Eq.~(\ref{eq:pmunuNLO}) is non-singular at the core of regions $R_1$ and $\bar{R}_1$.
The desired unpolarized structure functions are obtained by taking appropriate combinations of $g_{\mu \nu}$ and $P_\mu P_\nu$ as in Eq.~(\ref{eq:F1}).

\end{widetext}

\end{document}